\documentstyle[12pt]{article}

\newlength{\extralineskip}

\newcounter{equnum}[section]
\def\theequnum{\thesection.\arabic{equnum}}
\newcommand{\beq}{$$ \refstepcounter{equnum}}
\newcommand{\eeq}{\eqno (\theequnum) $$}
\newcommand{\bd}{\begin{displaymath}}
\newcommand{\ed}{\end{displaymath}}

\font\twlmsy=msbm10 at 12pt
\font\sevenmsy=msbm8
\font\fivemsy=msbm6
\newfam\Bbbfam
\textfont\Bbbfam=\twlmsy
\scriptfont\Bbbfam=\sevenmsy
\scriptscriptfont\Bbbfam=\fivemsy
\def\Bbb{\fam\Bbbfam}

\def\pvint{{\int\!\!\!\!\!\!-}}
\def\tr{{\rm tr}}
\def\e{~{\rm e}}
\makeatletter
\newdimen\normalarrayskip              
\newdimen\minarrayskip                 
\normalarrayskip\baselineskip
\minarrayskip\jot
\newif\ifold             \oldtrue            \def\new{\oldfalse}
\def\arraymode{\ifold\relax\else\displaystyle\fi} 
\def\@arrayskip{\ifold\baselineskip\z@\lineskip\z@
     \else
     \baselineskip\minarrayskip\lineskip2\minarrayskip\fi}
\def\@arrayclassz{\ifcase \@lastchclass \@acolampacol \or
\@ampacol \or \or \or \@addamp \or
   \@acolampacol \or \@firstampfalse \@acol \fi
\edef\@preamble{\@preamble
  \ifcase \@chnum
     \hfil$\relax\arraymode\@sharp$\hfil
     \or $\relax\arraymode\@sharp$\hfil
     \or \hfil$\relax\arraymode\@sharp$\fi}}
\def\@array[#1]#2{\setbox\@arstrutbox=\hbox{\vrule
     height\arraystretch \ht\strutbox
     depth\arraystretch \dp\strutbox
     width\z@}\@mkpream{#2}\edef\@preamble{\halign \noexpand\@halignto
\bgroup \tabskip\z@ \@arstrut \@preamble \tabskip\z@ \cr}%
\let\@startpbox\@@startpbox \let\@endpbox\@@endpbox
  \if #1t\vtop \else \if#1b\vbox \else \vcenter \fi\fi
  \bgroup \let\par\relax
  \let\@sharp##\let\protect\relax
  \@arrayskip\@preamble}

\begin{document}
\begin{titlepage}
\setcounter{footnote}0
\rightline{\baselineskip=12pt\vbox{\halign{&#\hfil\cr
&OUTP-96-56P\cr
&hep-th/9609237 &\cr
{   }&\cr &Revised Version &\cr {   }&\cr &\today\cr}}}
\vspace{0.5in}
\begin{center}
{\Large\bf Curvature Matrix Models for Dynamical Triangulations and the
Itzykson-Di~Francesco~Formula}\\
\medskip
\vskip0.5in
\baselineskip=12pt

\normalsize {\bf Richard J. Szabo}\footnote{\baselineskip=12pt Work supported
in part by the Natural Sciences and Engineering Research Council of
Canada.}$^,$\footnote{\baselineskip=12pt email address:
r.szabo1@physics.oxford.ac.uk} and {\bf John F.
Wheater}\footnote{\baselineskip=12pt email address:
j.wheater1@physics.oxford.ac.uk}
\medskip

\baselineskip=12pt

{\it Department of Physics, Theoretical Physics\\ University of Oxford\\ 1
Keble Road, Oxford OX1 3NP, U.K.}

\end{center}
\vskip1.0in

\begin{abstract}
\baselineskip=12pt

We study the large-$N$ limit of a class of matrix models for dually weighted
triangulated random surfaces using character expansion techniques. We show that
for various choices of the weights of vertices of the dynamical triangulation
the model can be solved by resumming the Itzykson-Di Francesco formula over
congruence classes of Young tableau weights modulo three. From this we show
that the large-$N$ limit implies a non-trivial correspondence with models of
random surfaces weighted with only even coordination number vertices. We
examine the critical behaviour and evaluation of observables and discuss their
interrelationships in all models. We obtain explicit solutions of the model for
simple choices of vertex weightings and use them to show how the matrix model
reproduces features of the random surface sum. We also discuss some general
properties of the large-$N$ character expansion approach as well as potential
physical applications of our results.

\end{abstract}
\end{titlepage}
\newpage

\setcounter{footnote}0
\baselineskip=18pt

\section{Introduction}

The statistical mechanics of random surfaces has been of much interest over the
years in the random geometry approach to two-dimensional quantum gravity and
lower dimensional string theory. These models can be solved non-perturbatively
by viewing discretized Riemann surfaces as Feynman graphs of $N\times N$ matrix
models (see \cite{fgz} for a review). The large-$N$ limit of the matrix model
exhibits phase transitions which correspond to the continuum limit of the
dynamically triangulated random surface model and whose large-$N$ expansion
coincides with the genus expansion of the string theory. In this paper we will
study a class of matrix models originally proposed by Das et al. \cite{das} in
which the weighting of the coordination numbers of vertices (describing
intrinsic curvature) of the dynamical triangulation of the surface can be
varied. These ``curvature" matrix models have been solved exactly in the
large-$N$ limit for a Penner interaction potential by Chekhov and Makeenko
\cite{chekmak}, and more recently for an arbitrary potential by Kazakov,
Staudacher and Wynter \cite{kaz1,kaz1a} using group character expansion
methods.

There are many problems of interest which could be solved from the large-$N$
limit of such matrix models. Foremost among these are questions related to the
quantum geometry and fractal structure of the surfaces which contribute to the
string partition function. For instance, the string theory ceases to make sense
for string embedding dimensions $D>1$. However, there is no obvious
pathological reason for the existence of a barrier in the statistical
mechanical model at $D=1$, and it has been suggested \cite{polymer} that the
statistical mechanical model evolves into a different geometrical phase for
$D>1$, for instance a branched polymer (tree-like) phase, rather than the
stringy (intrinsically two-dimensional) phase.  It was argued in \cite{das}
that curvature matrix models can represent such a transition by variation of
the weighting of the vertex coordination numbers. The trees are to be thought
of as connecting two-dimensional baby universes together in each of which the
usual $D<1$ behaviour is exhibited (see \cite{mak} for a review). Another
problem is the exact solution of two-dimensional quantum gravity with higher
curvature counterterms added to the action and the associated problem of the
existence of a phase transition from the gravitational phase to a flat phase of
the same string theory. By varying the vertex coordination numbers of a random
lattice to a flat one, the genus zero problem can be obtained as the
non-perturbative large-$N$ solution of an appropriate curvature matrix model
and it was demonstrated in \cite{kaz2} that there is no such phase transition.
Besides these problems, there are many other interesting physical situations
which require control over the types of surfaces that contribute to the random
sum.

The main subject of this paper is the curvature matrix model defined by the
partition function
\beq
Z_H[\lambda,t_q^*]=(2\pi\lambda)^{-N^2/2}\int
dX~\e^{-\frac{N}{2\lambda}~\tr~X^2+N~\tr~V_3(XA)}
\label{partfn}\eeq
where
\beq
V_3(XA)=\frac{1}{3}(XA)^3
\label{V3XA}\eeq
and the integration is over the space of $N\times N$ Hermitian matrices $X$
with $A$ an invertible $N\times N$ external matrix. The Feynman diagram
expansion of (\ref{partfn}) can be written symbolically as
\beq
Z_H[\lambda,t_q^*]=\sum_{G_3}\prod_{v_q^*\in
G_3}\left(\lambda^{q/2}t_q^*\right)^{{\cal N}(v_q^*)}
\label{partgraphs}\eeq
where the sum is over all fat-graphs $G_3$ made up of 3-point vertices, the
weight associated to a vertex $v_q^*$ of the lattice dual to $G_3$ of
coordination number $q$ is
\beq
t_q^*=\frac{1}{q}\frac{\tr}{N}A^q~~~~~,
\label{dualweights}\eeq
and ${\cal N}(v_q^*)$ is the number of such vertices in $G_3$. This matrix
model therefore assigns a weight $t_q^*$ whenever $q$ 3-point vertices bound a
face of the associated surface discretization, and it thus allows one to
control the local intrinsic curvature $R_q=\pi(6-q)/q$ of the dynamical
triangulation. We will examine the structure of the solution at large-$N$ when
these dual weights are arranged so that the only triangulations which
contribute to the sum in (\ref{partgraphs}) are those whose vertices have
coordination numbers which are multiples of three.

There are several technical and physical reasons for studying such a model. The
method of exact solution developed in \cite{kaz1,kaz1a} is based on a treatment
of the large-$N$ limit of the Itzykson-Di Francesco character expansion formula
\cite{di}. This formula is most naturally suited to matrix models of random
surfaces whose vertices have even coordination numbers, and therefore the
analysis in \cite{kaz1,kaz1a,kaz2} was restricted to such situations. In the
following we will show that the large-$N$ limit of the Itzykson-Di Francesco
formula for the curvature matrix model (\ref{partfn}) can be used provided that
one arranges the group character sum carefully. This arrangement reflects the
discrete symmetries of the triangulation model that are not present in the
models with vertices of only even coordination number. We shall see that this
solution of the curvature matrix model actually implies a graph theoretical
equivalence between the dynamically triangulated model and certain even
coordination number models which were studied in \cite{kaz1,kaz1a}. We will
also show how to map the Hermitian matrix model (\ref{partfn}) onto a complex
one whose character expansion is especially suited to deal with coordination
numbers which are multiples of three and whose observables are in a one-to-one
correspondence with those of the even coordination number matrix models. As a
specific example, we solve the model explicitly for a simple power law
variation of the vertex weights (\ref{dualweights}) which agrees with expected
results from the random surface sums and which provides explicit insights into
the non-perturbative behaviour of observables of the curvature matrix model
(\ref{partfn}), and also into the phase structure of the Itzykson-Zuber model
\cite{iz} which is related to higher-dimensional Hermitian matrix models
\cite{fgz,mak,semsz}. There are several physical situations which are best
described using such dynamical triangulations \cite{polymer}. The analysis
which follows also applies to quite general odd coordination number models and
establishes a non-trivial application of the techniques developed in
\cite{kaz1,kaz1a}.

The organization of this paper is as follows. In section 2 we derive the
character expansion of the matrix model (\ref{partfn}) and discuss the
structure of the Itzykson-Di Francesco formula in this case. We also show that
certain symmetry assumptions that are made agree with the mapping of the model
onto a complex curvature matrix model which demonstrates explicitly how the
character sum should be taken. In section 3 we discuss the large-$N$ saddle
point solutions of these matrix models. We show that the Itzykson-Di Francesco
formula implies non-trivial correspondences between the Feynman graphs of the
matrix model (\ref{partfn}) and those of some models of random surfaces with
even coordination numbers. We also establish this correspondence more directly
using graph theory arguments. We analyse the critical behaviour of the matrix
model when all vertices are weighted equally and discuss the various relations
that exist among observables of the matrix models. We also demonstrate
explicitly using the Wick expansion of (\ref{partfn}) that the large-$N$
solution of section 2 is indeed correct. In section 4 we examine a simple
situation where the dual vertices of the dynamical triangulation are not
weighted equally. We show how this affects the large-$N$ saddle-point solution
of the matrix model and thereby establish the validity of our solution for
generic choices of the vertex coupling constants. Our results here may also be
relevant to the study of phase transitions in the Itzykson-Zuber model
\cite{mak,semsz}. In section 5 we briefly discuss the difficult problems which
occur when dealing with matrix models that admit complex-valued saddle-points
of the Itzykson-Di Francesco formula, and section 6 contains some concluding
remarks and potential physical applications of our analysis.

\section{Large-$N$ Character Expansion of the Partition Function}

We begin by describing the character expansion method for solving the matrix
model (\ref{partfn}) in the large-$N$ limit \cite{kaz1,kaz1a}. The external
field $A$ in (\ref{partfn}) explicitly breaks the invariance of the model under
unitary transformations $X\to UXU^\dagger$ which diagonalize the Hermitian
matrix $X$. Thus, unlike the more conventional matrix models with $A={\bf1}$
for which the dual vertices $v_q^*$ are all weighted equally \cite{fgz}, the
partition function (\ref{partfn}) cannot be written as a statistical theory of
the eigenvalues of $X$. The number of degrees of freedom can still, however, be
reduced from $N^2$ to $N$ by expanding the invariant function
$\e^{\frac{N}{3}~\tr(XA)^3}$ in characters of the Lie group $GL(N,{\Bbb C})$ as
\cite{kaz1}
\beq
\e^{\frac{N}{3}~\tr(XA)^3}=c_N\sum_{\{h_i\}}{\cal X}_3[h]~\chi_{\{h_i\}}(XA)
\label{charexp}\eeq
where $c_N$ denotes an irrelevant numerical constant and
\beq
{\cal X}_3[h]=\left(\frac{N}{3}\right)^{\frac{1}{3}\sum_{i=1}^Nh_i}
\prod_{\epsilon=0,1,2}\frac{\Delta[h^{(\epsilon)}]}{\prod_{i=1}^{N/3}
\left(\frac{h_i^{(\epsilon)}-\epsilon}{3}\right)!}~{\rm sgn}\left[\prod_{0\leq
\epsilon_1<\epsilon_2\leq2}~\prod_{i,j=1}^{N/3}\left(h_i^{(\epsilon_2)}-h_j
^{(\epsilon_1)}\right)\right]
\label{char3}\eeq
The sum in (\ref{charexp}) is over unitary irreducible representations of
$GL(N,{\Bbb C})$. They are characterized by their Young tableau weights
$\{h_i\}_{i=1}^N$ which are increasing non-negative integers, $0\leq
h_i<h_{i+1}$, and are defined by $h_i=i-1+b_i$ where $b_i$ is the number of
boxes in row $i$ of the associated Young tableau. Because of the 3-valence
coupling on the left-hand side of (\ref{charexp}), the sum is supported on
those weights which can be factored into three groups of equal numbers of
integers $h^{(\epsilon)}_i$, $i=1,\dots,\frac{N}{3}$, where $\epsilon=0,1,2$
labels their congruence classes modulo 3. The $GL(N,{\Bbb C})$ characters can
be written using the Weyl character formula as
\beq
\chi_{\{h_i\}}(Y)=\frac{\det_{k,\ell}[y_k^{h_\ell}]}{\Delta[y]}
\label{weylchar}\eeq
where $y_i\in{\Bbb R}$ are the eigenvalues of the Hermitian matrix $Y$ and
\beq
\Delta[y]=\prod_{i<j}(y_i-y_j)=\det_{k,\ell}[y_k^{\ell-1}]
\eeq
is the Vandermonde determinant.

Substituting (\ref{charexp}) into (\ref{partfn}) and diagonalizing $X$, the
integration over unitary degrees of freedom can be carried out explicitly using
the Schur orthogonality relations for the characters. The resulting expression
is a statistical mechanics model in Young tableau weight space which is a
special case of the Itzykson-Di Francesco formula \cite{kaz1,di}
\beq
Z_H[\lambda,t_q^*]=c_N\sum_{h=\{h^e,h^o\}}\frac{\prod_{i=1}^{N/2}
(h_i^e-1)!!h_i^o!!}{\prod_{i,j=1}^{N/2}(h_i^e-h_j^o)}~{\cal
X}_3[h]~\chi_{\{h\}}
(A)\left(\frac{\lambda}{N}\right)^{-\frac{1}{4}N(N-1)+\frac{1}{2}\sum_{i=1}
^{N/2}(h_i^e+h_i^o)}
\label{diformula}\eeq
The sum in (\ref{diformula}) is restricted to the even representations of
$GL(N,{\Bbb C})$, i.e. those with Young tableau weights which split into an
equal number of even and odd integers $h_i^e$ and $h_i^o$,
$i=1,\dots,\frac{N}{2}$. This restriction arises because of the Gaussian
integration over the eigenvalues $x_i\in(-\infty,\infty)$ of the matrix $X$
\cite{kaz1}. Because of (\ref{char3}), these weights must also split equally
into their congruence classes modulo 3. The partition function
(\ref{diformula}) depends on only $N$ degrees of freedom (the Young tableau
weights $h_i$) and thus the curvature matrix model (\ref{partfn}) is formally
solvable in the large-$N$ limit.

For definiteness, we consider the matrix model of dually weighted discretized
surfaces for which the only graphs contributing in (\ref{partgraphs}) are those
triangulations which have $3m$, $m=1,2,\dots$, nearest-neighbour sites to each
vertex. This means that the vertex weights $t_q^*$ in (\ref{dualweights}) are
non-vanishing only when $q$ is a multiple of three. To  realize this explicitly
in the matrix model (\ref{partfn}), we take the external matrix $A$ to be of
the block form
\beq
A=A^{(3)}\equiv\pmatrix{\bar A^{1/3}&0&0\cr0&\omega_3\bar
A^{1/3}&0\cr0&0&\omega_3^2\bar A^{1/3}\cr}
\label{A3}\eeq
where $\omega_3\in{\Bbb Z}_3$ is a non-trivial cube root of unity and $\bar
A^{1/3}$ is an invertible $\frac{N}{3}\times\frac{N}{3}$ Hermitian matrix. The
character (\ref{weylchar}) can then be evaluated explicitly to get \cite{kaz1}
\beq
\chi_{\{h\}}(A^{(3)})=\chi_{\left\{\frac{h^{(0)}}{3}\right\}}(\bar
A)~\chi_{\left\{\frac{h^{(1)}-1}{3}\right\}}(\bar
A)~\chi_{\left\{\frac{h^{(2)}-2}{3}\right\}}(\bar A)~{\rm sgn}\left[
\prod_{0\leq\epsilon_1<\epsilon_2\leq2}~\prod_{i,j=1}^{N/3}\left(h_i^{
(\epsilon_2)}-h_j^{(\epsilon_1)}\right)\right]
\label{charA3}\eeq
and the statistical sum (\ref{diformula}) becomes
\beq\new{\begin{array}{c}
Z_H[\lambda,t_q^*]=c_N~\lambda^{-\frac14N(N-1)}\sum_{h=\{h^e,h^o\}}
\frac{\prod_i(h_i^e-1)!!h_i^o!!}
{\prod_{i,j}(h_i^e-h_j^o)}\prod_{\epsilon=0,1,2}\left(\frac{
\Delta[h^{(\epsilon)}]~\chi_{\left\{\frac{h^{(\epsilon)}-\epsilon}{3}
\right\}}(\bar A)}{\prod_i\left(\frac{h_i^{(\epsilon)}-\epsilon}{3}
\right)!}\right)\\~~~~~~~~~~~~~~~~
\times\e^{\sum_ih_i[\frac{1}{2}\log(\frac{\lambda}{N})+
\frac{1}{3}\log(\frac{N}{3})]}\end{array}}
\label{di3valence}\eeq
Notice that the sign factors from (\ref{char3}) and (\ref{charA3}) cancel each
other out in (\ref{di3valence}).

To treat the large-$N$ limit, we assume that at $N=\infty$ the sum over
representations in (\ref{di3valence}) becomes dominated by a single,
most-probable Young tableau $\{h_i\}$. The problem that immediately arises is
that the groupings of the Young tableau weights in (\ref{di3valence}) into
equal numbers of even and odd integers and into equal numbers of mod 3
congruence elements need not occur symmetrically. There does not appear to be
any canonical way to split the weights up and distribute them in the
appropriate way. However, it is natural to assume, by the symmetry of the
weight distributions in (\ref{di3valence}), that the saddle-point localizes
around that configuration with equal numbers of even and odd weights, {\it
each} set of which groups with equal numbers into their mod 3 congruence
classes $h_i^{e(\epsilon)}$ and $h_i^{o(\epsilon)}$, $i=1,\dots,\frac{N}{6}$.
It is also natural, by symmetry again, to further assume that although the
different sets of weights $h_i^{e(\epsilon)},h_i^{o(\epsilon)}$ do not
factorize and decouple from each other, the groupings of the weights into even
and odd mod 3 congruence classes are distributed in the same way. We therefore
assume this symmetrical splitting in (\ref{di3valence}) and write the
statistical sum over the mod 3 congruence classes of weights
$h_i^{(\epsilon)}$. We also make the additional assumption that the product of
weights from two different congruence class groupings contributes the same in
the large-$N$ limit as the Vandermonde determinant for a single grouping, i.e.
that
\beq
\prod_{i,j}\left(h_i^{(\epsilon_2)}-h_j^{(\epsilon_1)}\right)=\prod_{i\neq
j}\left(h_i^{(\epsilon_1)}-h_j^{(\epsilon_1)}\right)
\eeq
for $\epsilon_2\neq\epsilon_1$. As we shall see, these symmetrical grouping
assumptions are indeed valid and lead to the appropriate solution of the
curvature matrix model (\ref{partfn}) at large-$N$.

We now rescale $h_i\to N\cdot h_i$ in (\ref{di3valence}), simplify the
factorials for large-$N$ using the Stirling approximations $h!!\sim\e^{h(\log
h-1)/2}$ and $h!\sim\e^{(h+\frac{1}{2})\log h-h}$, and apply the above symmetry
assumption retaining only the leading planar (order $1/N$) contributions to
(\ref{di3valence}). After some algebra, the partition function
(\ref{di3valence}) for the symmetrically distributed weights $h^{(0)}$ in the
large-$N$ limit can be written as
\beq
Z_H^{(0)}\sim\sum_{h^{(0)}}\e^{N^2S_H[h^{(0)}]}
\label{partfnsym}\eeq
where the effective action is
\beq\new{\begin{array}{c}
S_H[h^{(0)}]=-\frac14\log\lambda+\frac{3}{2N^2}\sum_{i<j}^{N/3}
\log(h_i^{(0)}-h_j^{(0)})+\frac{1}{2N}
\sum_{i=1}^{N/3}h_i^{(0)}\left[2\log\lambda+\log\left(\lambda h_i^{(0)}
\right)-1\right]\\+
\frac{3}{N^2}\log I\left[\frac{Nh^{(0)}}{3},\bar A\right]\end{array}}
\label{effaction}\eeq
and we have introduced the Itzykson-Zuber integral \cite{iz}
\beq
I[h^{(0)},\bar
A]\equiv\int_{U(N/3)}dU~\e^{\sum_{i,j}h_i^{(0)}\alpha_j|U_{ij}|^2}=
\chi_{\{h^{(0)}\}}(\bar A)\frac{\Delta[a]}{\Delta[h^{(0)}]\Delta[\alpha]}
\label{izformula}\eeq
where $a_i$ are the eigenvalues of the matrix $\bar A$ and $\alpha_i\equiv\log
a_i$. In arriving at (\ref{partfnsym}) we have used the Vandermonde determinant
decomposition
\beq
\Delta[h^{(\epsilon)}]=\Delta[h^{e(\epsilon)}]\Delta[h^{o(\epsilon)}]
\prod_{i,j=1}^{N/6}\left(h_i^{e(\epsilon)}-h_j^{o(\epsilon)}\right)
\label{vandecompsym}\eeq
and ignored irrelevant terms which are independent of $\lambda$ and the $h$'s.

In the case of the even coordination number models studied in
\cite{kaz1,kaz1a,kaz2} the natural weight distribution to sum over in the
Itzykson-Di Francesco formula (\ref{diformula}) is that of the original even
representation restriction (see subsection 3.1). In our case the natural weight
distribution appears to be that of the mod 3 congruence classes. However, in
contrast to the even coordination number models, there is no way to justify
this from the onset. We shall see later on that the appropriate distribution
for the Itzykson-Di Francesco formula at large-$N$ is essentially determined by
the discrete symmetries of the given curvature matrix model. Some evidence for
the validity of this assumption comes from comparing the Hermitian model here
with the {\it complex} curvature matrix model
\beq
Z_C[\lambda,t_q^*]=(2\pi\lambda)^{-N^2}\int
d\phi~d\phi^\dagger~\e^{-\frac{N}{\lambda}~\tr~\phi^\dagger\phi-\frac{1}{3}~
\tr(\phi A\phi^\dagger B)^3}
\label{partfncomplex}\eeq
where the integration is now over the space of $N\times N$ complex-valued
matrices $\phi$, and $B$ is another external $N\times N$ invertible matrix.
Again one can expand the invariant, cubic interation term as in (\ref{charexp})
and the character expansion of (\ref{partfncomplex}) is \cite{kazproc}
\beq
Z_C[\lambda,t_q^*]=c_N\sum_{h=\{h^{(\epsilon)}\}}\frac{\prod_{i=1}^Nh_i!}
{\Delta[h]}~\chi_{\{h\}}(A)~\chi_{\{h\}}(B)~{\cal
X}_3[h]\left(\frac{\lambda}{N}
\right)^{-\frac{1}{2}N(N-1)+\sum_ih_i}
\label{complexcharexpB}\eeq
Now the Gaussian integration over the eigenvalues of the positive definite
Hermitian matrix $\phi^\dagger\phi$ goes over $[0,\infty)$ and so there is no
restriction to even and odd weights in (\ref{complexcharexpB}), only the
restriction to mod 3 congruence classes because of the 3-valence coupling. If
we take $B={\bf1}$, so that $\chi_{\{h\}}(B)\propto\Delta[h]$, and use the
decomposition (\ref{A3}), then the character expansion (\ref{complexcharexpB})
becomes
\beq
Z_C[\lambda,t_q^*]=c_N~\lambda^{-\frac12N(N-1)}
\sum_{h=\{h^{(\epsilon)}\}}\prod_{i=1}^Nh_i!
\prod_{\epsilon=0,1,2}\frac{\Delta[h^{(\epsilon)}]}{\prod_i\left(
\frac{h_i^{(\epsilon)}-\epsilon}{3}\right)!}~\chi_{\left\{\frac{h^{(\epsilon)}-
\epsilon}{3}\right\}}(\bar A)\e^{\sum_ih_i^{(\epsilon)}[\log(
\frac{\lambda}{N})+\frac{1}{3}\log(\frac{N}{3})]}
\label{ZCcharsplit}\eeq

Thus in this case the congruence classes of weights completely factorize, and
we can now naturally assume that the $N=\infty$ configuration of weights is
distributed equally into these three classes. In this sense, the complex matrix
model (\ref{partfncomplex}) is a better representative of the dually-weighted
triangulated random surface sum, because one need not make any ad-hoc
assumptions about the weight distribution. The character expansion
(\ref{ZCcharsplit}) suggests that the correct statistical distribution for the
dynamical triangulation model at large-$N$ is over $h_i^{(\epsilon)}$, as was
assumed above, and we shall discuss the explicit relationship between these
Hermitian and complex matrix models in the next section. We now rescale $h_i\to
N\cdot h_i$ and simplify (\ref{ZCcharsplit}) for large-$N$ as for the Hermitian
model. After some algebra, the partition function (\ref{ZCcharsplit}) for the
symmetrically distributed weights $h^{(0)}$ in the large-$N$ limit can be
written as
\beq
Z_C^{(0)}\sim\sum_{h^{(0)}}\e^{N^2S_C[h^{(0)}]}
\label{partfnsymC}\eeq
where the effective action is
\beq\new{\begin{array}{c}
S_C[h^{(0)}]=-\frac12\log\lambda+
\frac{6}{N^2}\sum_{i<j}^{N/3}\log(h_i^{(0)}-h_j^{(0)})+\frac{2}{N}
\sum_{i=1}^{N/3}h_i^{(0)}\left[\log\left(\lambda^{3/2}h_i^{(0)}
\right)-1\right]\\
+\frac{3}{N^2}\log I\left[\frac{Nh^{(0)}}{3},\bar A\right]\end{array}}
\label{effactionC}\eeq

\section{Saddle-point Solutions}

If the characteristic $h$'s which contribute to the sum in (\ref{partfnsym})
are of order $1$, then the large-$N$ limit of the Itzykson-Di Francesco formula
can be found using the saddle-point approximation. The attraction or repulsion
of the $\lambda$-dependent terms in (\ref{di3valence}) is compensated by the
Vandermonde determinant factors, so that, in spite of the unsymmetrical
grouping of Young tableau weights that occurs, the partition function is
particularly well-suited to a usual saddle-point analysis in the large-$N$
limit. In this section we shall see that the symmetry assumptions made above
give a well-posed solution of the matrix model at $N=\infty$. The partition
functions (\ref{partfn}) and (\ref{partfncomplex}) are dominated at large-$N$
by their saddle points which are the extrema of the actions in
(\ref{effaction}) and (\ref{effactionC}). In what follows we shall determine
the saddle-point solutions of both the Hermitian and complex matrix models and
use them to examine various properties of the random surface sum.

\subsection{Hermitian Model}

To find the saddle-point solution of the Hermitian matrix model, we minimize
the action (\ref{effaction}) with respect to the Young tableau weights. Then
the stationary condition $\frac{\partial S_H}{\partial h_i^{(0)}}=0$ leads to
the saddle-point equation
\beq
\frac{3}{N}\sum_{\buildrel{j=1}\over{j\neq
i}}^{N/3}\frac{1}{h_i^{(0)}-h_j^{(0)}}=-\log(\lambda^3h_i^{(0)})-2{\cal
F}(h_i^{(0)})
\label{saddleHdiscr}\eeq
where we have introduced the Itzykson-Zuber correlator
\beq
{\cal F}(h_i^{(0)})=3\frac{\partial\log I[h^{(0)}/3,\bar A]}{\partial
h_i^{(0)}}
\label{izcorr}\eeq

To solve (\ref{saddleHdiscr}), we assume that at large-$N$ the Young tableau
weights $h_i^{(0)}$, $i=1,\dots,\frac{N}{3}$, become distributed on a finite
interval $h\in[0,a]$ on the real line and introduce the normalized spectral
density $\rho_H(h)\equiv\frac{dx}{dh(x)}$, where $h(x)\in[0,a]$ is a
non-decreasing differentiable function of $x\in[0,1]$ with $h(3i/N)=N\cdot
h_i^{(0)}$. Discrete sums over $h_j^{(0)}$ are then replaced with integrals
over $h$ by the rule $\frac{3}{N}\sum_{j=1}^{N/3}\to\int_0^a dh$. Notice that
since the $h_i$'s are increasing integers, we have $0\leq\rho_H(h)\leq1$, and
so the spectral distribution is trivial, $\rho_H(h)=1$, on some sub-interval
$[0,b]$ with $0<b\leq1\leq a$ \cite{dougkaz}. The saddle-point equation
(\ref{saddleHdiscr}) then becomes an integral equation for the spectral
density,
\beq
\pvint_{\!\!b}^a~dh'~\frac{\rho_H(h')}{h-h'}=-\log(\lambda^3h)-
\log\left(\frac{h}{h-b}\right)-2{\cal F}(h)~~~~~,~~~~~h\in[b,a]
\label{saddlepteq}\eeq
where we have saturated the spectral distribution function at its maximum value
$\rho_H(h)=1$ on $[0,b]$. The saddle-point solution of the matrix model can
thus be determined as the solution of the Riemann-Hilbert problem for the usual
resolvent function
\beq
{\cal H}_H(h)=\left\langle\frac{3}{N}\sum_{i=1}^{N/3}\frac{1}{h-h_i^{(0)}}
\right\rangle=\int_0^adh'~\frac{\rho_H(h')}{h-h'}
\label{resolv}\eeq
which is analytic everywhere in the complex $h$-plane away from the support
interval $[0,a]$ of $\rho_H$ where it has a branch cut. The discontinuity of
the resolvent across this cut determines the spectral density by
\beq
{\cal H}_H(h\pm i0)=\pvint_{\!\!0}^a~dh'~\frac{\rho_H(h')}{h-h'}\mp
i\pi\rho_H(h)~~~,~~~h\in[0,a]
\label{disceq}\eeq

In contrast to the more conventional Hermitian one-matrix models \cite{fgz},
the Riemann-Hilbert problem (\ref{saddlepteq}) involves the unknown
Itzykson-Zuber correlator ${\cal F}(h)$ which must be determined separately in
the large-$N$ limit. As shown in \cite{kaz1,kaz1a}, it is determined by the
vertex couplings (\ref{dualweights}) through the contour integral
\beq
q\tilde t_q^*\equiv3qt_{3q}^*=3\frac{\tr}{N}\bar A^q=\frac{1}{q}\oint_{\cal
C}\frac{dh}{2\pi i}~\e^{q({\cal H}_H(h)+{\cal F}(h))}~~~,~~~q\geq1
\label{weightcont}\eeq
where the closed contour $\cal C$ encircles the support of the spectral
function $\rho_H(h)$, i.e. the cut singularity of ${\cal H}_H(h)$, with
counterclockwise orientation in the complex $h$-plane. Note that
(\ref{weightcont}) is the large-$N$ limit of a set of weights of size $N/3$ and
it follows from the identity \cite{kaz1}
\beq
\tr~\bar A^q=\sum_{k=1}^{N/3}\frac{\chi_{\left\{\tilde
h^{(0)}_k(q)/3\right\}}(\bar A)}{\chi_{\left\{h^{(0)}/3\right\}}(\bar A)}
\label{weightNid}\eeq
where
\beq
(\tilde h^{(0)}_k(q))_i=h^{(0)}_i+3q\delta_{ik}
\label{tildehk}\eeq
In the next section we shall discuss the evaluation of ${\cal F}(h)$ and the
corresponding structure of the curvature matrix model as the vertex weights
$\tilde t_q^*$ are varied. Notice that, strictly speaking, in most cases of
interest the characters corresponding to a specific choice of vertex weightings
cannot be represented via matrix traces as in (\ref{dualweights}) and need to
be defined by an analytical continuation. This can be accomplished by using the
Schur-Weyl duality theorem to represent the $GL(N,{\Bbb C})$ characters as the
generalized Schur functions
\beq
\chi_{\{h\}}[t^*]=\det_{k,\ell}\left[P_{h_k+1-\ell}[t^*]\right]
\label{schurchar}\eeq
where $P_n[t^*]$ are the Schur polynomials defined by
\beq
\exp\left(N\sum_{q=1}^\infty z^qt_q^*\right)=\sum_{n=0}^\infty z^nP_n[t^*]
\label{schurpoly}\eeq
When the weights $t_q^*$ are given by (\ref{dualweights}), the Schur functions
(\ref{schurchar}) coincide with the Weyl characters (\ref{weylchar}).

Our first observation here is that the saddle-point equation (\ref{saddlepteq})
is identical to that of the even-even coordination number model discussed in
\cite{kaz1,kaz1a} which is defined by replacing the cubic interaction matrix
potential $V_3(XA)$ in (\ref{partfn}) by
\beq
V_{\rm even}(XA)=\sum_{q=1}^\infty t_{2q}(XA^{(2)})^{2q}
\label{VevenXA}\eeq
where
\beq
A^{(2)}=\pmatrix{\tilde A^{1/2}&0\cr0&-\tilde A^{1/2}\cr}
\label{Aeven}\eeq
with $\tilde A^{1/2}$ an invertible $\frac N2\times\frac N2$ Hermitian matrix,
and
\beq
2qt_{2q}=2qt_{2q}^*=2\frac{\tr}{N}\tilde A^q
\label{t2qdef}\eeq
The curvature matrix model with potential (\ref{VevenXA}) generates a sum over
random surfaces where arbitrary even coordination number vertices of the
primary and dual lattices are permitted and are weighted identically, so that
the model (\ref{VevenXA}) is self-dual. In fact, the same saddle-point equation
arises from the interaction potential
\beq
V_4(XA)=\frac{1}{4}(XA^{(2)})^4
\label{V4}\eeq
The matrix model with potential $V_4$ generates discretizations built up from
even-sided polygons with 4-point vertices.

As shown in \cite{kaz1}, the Itzykson-Di Francesco formula for the curvature
matrix model with potential (\ref{VevenXA}) follows from replacing ${\cal
X}_3[h]$ in (\ref{diformula}) by
\beq
\chi_{\{h\}}(A^{(2)})=\chi_{\left\{\frac{h^e}2\right\}}(\tilde
A)~\chi_{\left\{\frac{h^o-1}{2}\right\}}(\tilde A)~{\rm
sgn}\left[\prod_{i,j=1}^{N/2}\left(h_i^e-h_j^o\right)\right]
\label{chiAeven}\eeq
Now the even and odd weights can be naturally assumed to distribute equally,
and the effective action is
\beq
S_{\rm even}[h^e]=-\frac14\log\lambda
+\frac2{N^2}\sum_{i<j}^{N/2}\log(h_i^e-h_j^e)+\frac1N
\sum_{i=1}^{N/2}h_i^e\left[\log(\lambda h_i^e)-1\right]+4\log
I\left[\frac{Nh^e}2,\tilde A\right]
\label{Seven}\eeq
Defining a distribution function for the $N/2$ weights $h_i^e$ analogously to
that above and varying the action (\ref{Seven}) yields the saddle-point
equation (\ref{saddlepteq}) with $\lambda^3\to\lambda$. In this case the
Itzykson-Zuber correlator ${\cal F}(h)$ is determined as in (\ref{weightcont})
but now with $q\tilde t_q^*$ defined to be equal to (\ref{t2qdef}) (i.e. the
large-$N$ limit of a set of weights of size $N/2$).

Likewise, in the case of the 4-point model (\ref{V4}), we replace ${\cal
X}_3[h]$ by
\beq
{\cal X}_4[h]=\left(\frac{N}{4}\right)^{\frac14\sum_ih_i}\prod_{\mu=0}^3
\frac{\Delta[h^{(\mu)}]}{\prod_{i=1}^{N/4}\left(\frac{h_i^{(\mu)}-
\mu}4\right)!}~{\rm
sgn}\left[\prod_{0\leq\mu_1<\mu_2\leq3}\prod_{i,j=1}^{N/4}\left
(h_i^{(\mu_2)}-h_j^{(\mu_1)}\right)\right]
\label{X4h}\eeq
where now the character sum is supported on weights that factor equally into
their congruence classes $h_i^{(\mu)}$, $\mu=0,\dots,4$, $i=1,\dots,\frac N4$,
modulo 4. Because of the original weight constraint of the Itzykson-Di
Francesco formula, this means that the even weights $h_i^e$ distribute equally
into the mod 4 congruence classes $\mu=0,2$ and the odd weights $h_i^o$ into
the classes $\mu=1,3$. Again, assuming these weights all distribute equally
leads to the effective action
\beq
S_4[h^e]=-\frac14\log\lambda+\frac1{N^2}\sum_{i<j}^{N/2}
\log(h_i^e-h_j^e)+\frac1{2N}\sum_{i=1}^{N/2}h_i^e\left[\log(\lambda^2
h_i^e)-1\right]+2\log I\left[\frac{Nh^e}{2},\tilde A\right]
\label{S4}\eeq
Introducing a distribution function for the $N/2$ weights $h_i^e$ again leads
to precisely the same saddle-point equation (\ref{saddlepteq}) with
$\lambda^3\to\lambda^2$. Here and in the even-even model above, unlike the
3-point model of the previous section, the large-$N$ configuration of weights
naturally localizes onto $h_i^e$.

These three matrix models therefore all possess the same solution at
$N=\infty$, i.e. their random surface ensembles of genus zero graphs are
identical. Their $1/N$ corrections will differ somewhat because of the
different ways that the weights split into the respective congruence classes in
the three separate cases. The genus zero free energies are related in a simple
fashion. From (\ref{effaction}) and the definition of the Hermitian
distribution function, the genus zero free energy for the 3-point model is
\beq\new{\begin{array}{c}
S_H[\lambda,\tilde
t_q^*]=\frac{1}{6}\left(\frac{1}{2}\int_0^a\!\!\pvint_{\!\!0}^adh~dh'~\rho_H
(h)\rho_H(h')\log|h-h'|+\int_0^adh~\rho_H(h)h\left[\log(\lambda^3h)-1\right]
\right.\\\left.+2\int_0^adh~\rho_H(h)\log I_c[h,\bar
A]\right)-\frac14\log\lambda\end{array}}
\label{SHsaddlept}\eeq
where $\rho_H(h)$ is the solution of the saddle-point equation
(\ref{saddlepteq}) and $\log I_c[h,\bar A]=3^2\log I[\frac{Nh^{(0)}}3,\bar A]$
is the Itzykson-Zuber integral (\ref{izformula}) at $N=\infty$. Similarly, the
genus zero free energies for the even-even and 4-point models can be written
using the same distribution function $\rho_H(h)$. Now, however, the rules for
replacing sums by integrals differ. For the even-even and 4-point models the
normalized sums corresponding to a spectral density normalized to unity are
$\frac2N\sum_{i=1}^{N/2}$ and the $N=\infty$ Itzykson-Zuber integral is $\log
I_c[h,\tilde A]=2^2\log I[\frac{Nh^e}2,\tilde A]$. Taking this into account
along with the change of $\lambda$ in the three cases, we find that the free
energies of the three models discussed above are all related by
\beq
3S_H[\lambda^{2/3},\tilde t_q^*]=2S_4[\lambda,\tilde t_q^*]=S_{\rm
even}[\lambda^2,\tilde t_q^*]
\label{freerels}\eeq

It is quite intriguing that the Itzykson-Di Francesco formula naturally implies
these relations by dictating the manner in which the Young tableau weights
should decompose with respect to the appropriate congruence classes in each
case. This is reflected in both the overall numerical coefficients and the
overall powers of $\lambda$ that appear in (\ref{freerels}). In the next
subsection we shall give purely graph-theoretical proofs of these
relationships. This yields a non-trivial verification of the assumptions made
in section 2 about the splitting of the Young tableau weights for the dynamical
triangulation model.

\subsection{Graph Theoretical Relationships}

Before pursuing further the properties of the curvature matrix model above, we
present a direct, graphical proof of the relationship (\ref{freerels}) between
the generating functions for the 3-point, 4-point, and even-even planar graphs.
We denote the ensembles of spherical graphs in these three cases by ${\cal
M}_3$, ${\cal M}_4$ and ${\cal M}_{\rm even}$, respectively. We first discuss
the mapping ${\cal M}_3\to{\cal M}_4$. Consider a planar graph $G\in{\cal
M}_3$. Using a version of the colouring theorem for a spherical topology, we
can colour the lines of $G$ by three colours, R, B, and G, to give a planar
graph $G^c$ with labelled lines (Fig. 1). Because $G$ consists of polygons
whose numbers of sides are all multiples of three, the colouring can be chosen
so that the three colours always occur in the sequence (R,B,G) as one goes
around any polygon of $G^c$ with clockwise orientation. We can now contract the
two 3-point vertices bounding each R-link of $G^c$ into a 4-point vertex (Fig.
2). A polygon of $3m$ sides with 3-point vertices in $G^c$ then becomes a
polygon of $2m$ sides with 4-point vertices. The resulting 2-coloured graph
$\tilde G^c$ thus belongs to ${\cal M}_4$. Notice that if each link has a
weight $\lambda$ associated to it, then the effect of the R-contractions is to
map $\lambda^3\to\lambda^2$ on ${\cal M}_3\to{\cal M}_4$.

\begin{figure}
\unitlength=0.90mm
\linethickness{0.4pt}
\begin{picture}(150.00,90.00)(0,10)
\small
\thicklines
\put(70.00,35.00){\line(1,0){30}}
\put(70.00,35.00){\line(-1,2){10}}
\put(60.00,55.00){\line(1,2){10}}
\put(100.00,35.00){\line(1,2){10}}
\put(110.00,55.00){\line(-1,2){10}}
\put(70.00,75.00){\line(1,0){30}}
\put(70.00,35.00){\line(-1,-1){10}}
\put(70.00,75.00){\line(-1,1){10}}
\put(100.00,35.00){\line(1,-1){10}}
\put(100.00,75.00){\line(1,1){10}}
\put(60.00,55.00){\line(-1,0){14}}
\put(110.00,55.00){\line(1,0){14}}
\put(110.00,25.00){\line(1,-2){5}}
\put(110.00,25.00){\line(1,0){11}}
\put(60.00,85.00){\line(-1,2){5}}
\put(60.00,85.00){\line(-1,0){11}}
\thinlines
\put(85.00,55.00){\line(0,1){40}}
\put(87.00,77.00){\makebox(0,0){R}}
\put(85.00,55.00){\line(0,-1){40}}
\put(87.00,37.00){\makebox(0,0){R}}
\put(80.00,17.50){\line(3,2){40}}
\put(90.00,17.50){\line(-3,2){40}}
\put(85.00,55.00){\line(2,-1){35}}
\put(85.00,55.00){\line(-2,-1){35}}
\put(85.00,55.00){\line(2,1){35}}
\put(105.00,82.50){\makebox(0,0){G}}
\put(65.00,82.50){\makebox(0,0){B}}
\put(107.00,68.00){\makebox(0,0){B}}
\put(117.00,57.00){\makebox(0,0){R}}
\put(107.00,43.00){\makebox(0,0){G}}
\put(63.00,43.00){\makebox(0,0){B}}
\put(63.00,68.00){\makebox(0,0){G}}
\put(105.00,27.50){\makebox(0,0){B}}
\put(65.00,27.50){\makebox(0,0){G}}
\put(85.00,55.00){\line(-2,1){35}}
\put(80.00,95.00){\line(3,-2){40}}
\put(90.00,95.00){\line(-3,-2){40}}
\put(115.00,17.50){\line(0,1){60}}
\put(117.00,57.00){\makebox(0,0){R}}
\put(55.00,35.00){\line(0,1){57.5}}
\put(53.00,57.00){\makebox(0,0){R}}
\put(80.00,21.00){\line(1,0){40}}
\put(90.00,91.00){\line(-1,0){40}}
\put(53.00,83.00){\makebox(0,0){R}}
\put(60.00,92.00){\makebox(0,0){G}}
\put(117.00,27.00){\makebox(0,0){R}}
\put(109.00,22.00){\makebox(0,0){G}}
\end{picture}
\begin{description}
\small
\baselineskip=12pt
\item[Figure 1:] A 3-colouring of a graph in ${\cal M}_3$ (thick lines) and its
associated dual graph in ${\cal M}_3^*$ (thin lines).
\end{description}
\end{figure}

\begin{figure}
\unitlength=0.90mm
\linethickness{0.4pt}
\begin{picture}(150.00,90.00)(0,10)
\small
\thicklines
\put(85.00,35.00){\line(1,1){20}}
\put(85.00,35.00){\line(-1,1){20}}
\put(85.00,75.00){\line(-1,-1){20}}
\put(85.00,75.00){\line(1,-1){20}}
\put(85.00,35.00){\circle*{2.00}}
\put(85.00,75.00){\circle*{2.00}}
\put(85.00,35.00){\line(-1,-1){10}}
\put(85.00,35.00){\line(1,-1){10}}
\put(65.00,55.00){\circle*{2.00}}
\put(105.00,55.00){\circle*{2.00}}
\put(95.00,25.00){\circle*{2.00}}
\put(95.00,25.00){\line(-1,-1){6}}
\put(95.00,25.00){\line(1,1){6}}
\put(95.00,25.00){\line(1,-1){6}}
\put(65.00,55.00){\line(-1,1){10}}
\put(65.00,55.00){\line(-1,-1){10}}
\put(85.00,75.00){\line(1,1){10}}
\put(85.00,75.00){\line(-1,1){10}}
\put(75.00,85.00){\circle*{2.00}}
\put(75.00,85.00){\line(-1,1){6}}
\put(75.00,85.00){\line(-1,-1){6}}
\put(75.00,85.00){\line(1,1){6}}
\put(105.00,55.00){\line(1,1){10}}
\put(105.00,55.00){\line(1,-1){10}}
\thinlines
\put(85.00,55.00){\line(1,1){20}}
\put(85.00,55.00){\line(-1,-1){20}}
\put(85.00,55.00){\line(-1,1){25}}
\put(85.00,55.00){\line(1,-1){25}}
\put(100.00,70.00){\line(-1,1){25}}
\put(100.00,70.00){\line(1,-1){20}}
\put(70.00,70.00){\line(1,1){20}}
\put(70.00,70.00){\line(-1,-1){20}}
\put(70.00,40.00){\line(-1,1){20}}
\put(70.00,40.00){\line(1,-1){25}}
\put(100.00,40.00){\line(1,1){20}}
\put(100.00,40.00){\line(-1,-1){20}}
\put(80.00,96.00){\line(-1,-1){20}}
\put(98.00,22.00){\line(1,1){15}}
\put(98.00,22.00){\line(-1,-1){10}}
\multiput(85.00,55.00)(0,2){20}{\line(0,1){1}}
\multiput(85.00,55.00)(0,-2){20}{\line(0,-1){1}}
\multiput(85.00,55.00)(2,0){20}{\line(1,0){1}}
\multiput(85.00,55.00)(-2,0){20}{\line(-1,0){1}}
\multiput(85.00,85.00)(2,0){7}{\line(1,0){1}}
\multiput(85.00,85.00)(-2,0){5}{\line(-1,0){1}}
\multiput(75.00,85.00)(-2,-1){10}{\line(3,0){1}}
\multiput(85.00,25.00)(-2,0){7}{\line(-1,0){1}}
\multiput(85.00,25.00)(2,0){5}{\line(1,0){1}}
\multiput(95.00,25.00)(2,1){10}{\line(3,0){1}}
\multiput(115.00,55.00)(0,2){8}{\line(0,1){1}}
\multiput(115.00,55.00)(0,-2){8}{\line(0,-1){1}}
\multiput(55.00,55.00)(0,2){6}{\line(0,1){1}}
\multiput(55.00,55.00)(0,-2){6}{\line(0,-1){1}}
\multiput(62.00,78.00)(0,2){6}{\line(0,1){1}}
\multiput(62.00,78.00)(0,-2){6}{\line(0,-1){1}}
\multiput(108.00,32.00)(0,2){5}{\line(0,1){1}}
\multiput(108.00,32.00)(0,-2){5}{\line(0,-1){1}}
\put(77.00,90.00){\makebox(0,0){G}}
\put(83.00,80.00){\makebox(0,0){B}}
\put(87.00,80.00){\makebox(0,0){G}}
\put(95.00,68.00){\makebox(0,0){B}}
\put(75.00,68.00){\makebox(0,0){G}}
\put(94.00,48.00){\makebox(0,0){G}}
\put(76.00,48.00){\makebox(0,0){B}}
\put(79.00,32.00){\makebox(0,0){G}}
\put(91.00,32.00){\makebox(0,0){B}}
\put(89.00,22.00){\makebox(0,0){G}}
\end{picture}
\begin{description}
\small
\baselineskip=12pt
\item[Figure 2:] A 2-colouring of a graph in ${\cal M}_4$ (thick lines), its
associated dual graph in ${\cal M}_4^*$ (thin lines), and the corresponding
diagram in ${\cal M}_{\rm even}$ (dashed lines). The 4-point vertices denoted
by solid circles are contracted from the R-links of the corresponding graph of
${\cal M}_3$ depicted in Fig. 1.
\end{description}
\end{figure}

Conversely, suppose $\tilde G\in{\cal M}_4$. We can form a 2-coloured graph
$\tilde G^c$, with colours B and G, such that the colours alternate along the
lines of $\tilde G^c$ (Fig. 2). There are two possible ways to orient lines of
$\tilde G^c$, by defining a direction to them from either B to G or from G to B
at each vertex. The 4-point vertices of the resulting oriented graph can then
be split into a line, labelled by R, bounded by two 3-point vertices. There are
two ways of doing this, by splitting the 4-point vertex either vertically or
horizontally with respect to the given orientation (Fig. 3). Thus to each
2-coloured graph $\tilde G^c\in{\cal M}_4$ there corresponds two distinct
(topologically inequivalent) 3-coloured graphs $G^c\in{\cal M}_3$. There are
also three distinct 2-coloured graphs $\tilde G^c\in{\cal M}_4$ for each
3-coloured graph $G^c\in{\cal M}_3$ corresponding to the three possible choices
of contraction colour R, B or G. This therefore defines a three-to-two mapping
on ${\cal M}_4\to{\cal M}_3$ and is just the statement of the first equality of
(\ref{freerels}).

\begin{figure}
\unitlength=0.90mm
\linethickness{0.4pt}
\begin{picture}(150.00,50.00)(0,10)
\small
\thicklines
\put(10.00,35.00){\circle*{2.00}}
\put(110.00,35.00){\circle*{2.00}}
\put(0.00,45.00){\line(1,-1){20}}
\put(0.00,25.00){\line(1,1){20}}
\put(0.00,45.00){\makebox(0,0){\Large$\searrow$}}
\put(0.00,25.00){\makebox(0,0){\Large$\nearrow$}}
\put(15.00,40.00){\makebox(0,0){\Large$\nearrow$}}
\put(15.00,30.00){\makebox(0,0){\Large$\searrow$}}
\put(30.00,35.00){\makebox(0,0){$\longrightarrow$}}
\put(50.00,42.00){\line(0,-1){14}}
\put(50.00,42.00){\line(-1,1){10}}
\put(50.00,42.00){\line(1,1){10}}
\put(50.00,28.00){\line(1,-1){10}}
\put(50.00,28.00){\line(-1,-1){10}}
\put(40.00,52.00){\makebox(0,0){\Large$\searrow$}}
\put(55.00,47.00){\makebox(0,0){\Large$\nearrow$}}
\put(40.00,18.00){\makebox(0,0){\Large$\nearrow$}}
\put(55.00,23.00){\makebox(0,0){\Large$\searrow$}}
\put(100.00,45.00){\line(1,-1){20}}
\put(100.00,25.00){\line(1,1){20}}
\put(100.00,45.00){\makebox(0,0){\Large$\searrow$}}
\put(105.00,30.00){\makebox(0,0){\Large$\swarrow$}}
\put(120.00,45.00){\makebox(0,0){\Large$\swarrow$}}
\put(115.00,30.00){\makebox(0,0){\Large$\searrow$}}
\put(130.00,35.00){\makebox(0,0){$\longrightarrow$}}
\put(150.00,35.00){\line(1,0){14}}
\put(150.00,35.00){\line(-1,1){10}}
\put(150.00,35.00){\line(-1,-1){10}}
\put(164.00,35.00){\line(1,1){10}}
\put(164.00,35.00){\line(1,-1){10}}
\put(140.00,45.00){\makebox(0,0){\Large$\searrow$}}
\put(145.00,30.00){\makebox(0,0){\Large$\swarrow$}}
\put(174.00,45.00){\makebox(0,0){\Large$\swarrow$}}
\put(169.00,30.00){\makebox(0,0){\Large$\searrow$}}
\put(0.00,47.00){\makebox(0,0){B}}
\put(20.00,47.00){\makebox(0,0){G}}
\put(0.00,23.00){\makebox(0,0){G}}
\put(20.00,23.00){\makebox(0,0){B}}
\put(60.00,54.00){\makebox(0,0){G}}
\put(40.00,54.00){\makebox(0,0){B}}
\put(60.00,16.00){\makebox(0,0){B}}
\put(40.00,16.00){\makebox(0,0){G}}
\put(52.00,35.00){\makebox(0,0){R}}
\put(157.00,37.00){\makebox(0,0){R}}
\put(120.00,47.00){\makebox(0,0){G}}
\put(100.00,47.00){\makebox(0,0){B}}
\put(100.00,23.00){\makebox(0,0){G}}
\put(120.00,23.00){\makebox(0,0){B}}
\put(140.00,23.00){\makebox(0,0){G}}
\put(174.00,47.00){\makebox(0,0){G}}
\put(140.00,47.00){\makebox(0,0){B}}
\put(174.00,23.00){\makebox(0,0){B}}
\end{picture}
\begin{description}
\small
\baselineskip=12pt
\item[Figure 3:] The two possible 4-point vertex splittings which respect the
given B and G colour orientation.
\end{description}
\end{figure}

Actually, there exists a simpler, line mapping between these two ensembles of
graphs in terms of dual lattices. Let ${\cal M}_3^*$ denote the collection of
planar graphs dual to those of ${\cal M}_3$ (i.e. lattices built up from
triangles that form $3m$-valence vertices), and ${\cal M}_4^*$ the dual
ensemble to ${\cal M}_4$ (i.e. the graphs formed of squares that meet to form
vertices of even coordination number). The sets ${\cal M}_3^*$ and ${\cal
M}_4^*$ are generated, respectively, by the curvature matrix models with matrix
potentials
\beq
V_3^{(*)}(XA)=\sum_{q=1}^\infty
t_{3q}^*(XA_3)^{3q}~~~~~,~~~~~V_4^{(*)}(XA)=\sum_{q=1}^\infty
t_{2q}(XA_4)^{2q}
\label{V3*}\eeq
where the $N\times N$ matrix $A_m$ is defined by $\frac\tr NA_m^k=\delta^k_m$.
The corresponding generating functions $S_H^{(*)}[\lambda,\tilde t_q^*]$ and
$S_4^{(*)}[\lambda,\tilde t_q^*]$ coincide, respectively, with
(\ref{effaction}) and (\ref{S4}) \cite{kaz1}. Again the lines of the graphs of
${\cal M}_3^*$ can be 3-coloured, and the deletion of all R-coloured lines
gives a map ${\cal M}_3^*\to{\cal M}_4^*$ (Figs. 1 and 2). This mapping is
three-to-one because of the three possible choices of deletion colour. The
inverse map ${\cal M}_4^*\to{\cal M}_3^*$, defined by inserting an R-link
across the diagonal of each square of the graphs of ${\cal M}_4^*$, is then
two-to-one because of the two possible choices of diagonal of a square.

The correspondence between the ensembles of graphs ${\cal M}_4$ and ${\cal
M}_{\rm even}$ is similar and has been noted in \cite{kaz1a} (see Fig. 2).
Given a graph $G^*\in{\cal M}_4^*$, we choose a diagonal of each square of
$G^*$. Connecting these diagonals together produces a graph in ${\cal M}_{\rm
even}$. Equivalently, we can place vertices in the face centers of the
corresponding dual graph $G\in{\cal M}_4$ and connect them together by drawing
lines through each of the 4-point vertices of $G$ (so that
$\lambda^4\to\lambda^8$ from the splitting of these 4-point vertices). In this
way we obtain a map ${\cal M}_4,{\cal M}_4^*\to{\cal M}_{\rm even}$ where the
vertices and face centers of graphs of ${\cal M}_{\rm even}$ correspond to the
vertices of graphs of ${\cal M}_4^*$, or equivalently the faces of ${\cal
M}_4$. Because there are two distinct ways of choosing the diagonal of a square
of $G^*\in{\cal M}_4^*$ (equivalently two ways of splitting a 4-point vertex of
$G\in{\cal M}_4$ with respect to a 2-colour orientation analogously to that in
Fig. 3), this mapping is two-to-one and corresponds to the second equality of
(\ref{freerels}). In particular, there is a three-to-one correspondence between
graphs of ${\cal M}_3,{\cal M}_3^*$ and ${\cal M}_{\rm even}$.

We stress that these graphical mappings are only valid on the sphere. In terms
of the Itzykson-Di Francesco formula, this means that the ${\cal O}(1/N)$
corrections to the large-$N$ saddle-point solutions of the corresponding
curvature matrix models will differ. These non-trivial correspondences are
predicted from the matrix model formulations, because in all cases we obtain
the same $N=\infty$ saddle-point equations but different splittings of Young
tableau weights into congruence classes thus leading to different overall
combinatorical factors in front of the graph generating function. Note that in
the case of the 4-point and even-even models the vertex weights $\tilde t_q^*$
are mapped into each other under the above graphical correspondence because of
the simple line map that exists between ${\cal M}_{\rm even}$ and ${\cal
M}_4^*$. In the case of the mappings onto ${\cal M}_3$ the weights $\tilde
t_q^*$ defined in (\ref{weightcont}) are mapped onto those defined by
(\ref{t2qdef}) because of the contraction of $3m$-sided polygons into
$2m$-sided polygons.

\subsection{Complex Model}

As mentioned in section 2, it is useful to compare the Hermitian 3-point
curvature matrix model with the complex one since in the latter case the Young
tableau weights completely factorize and the splitting of them into mod 3
congruence classes appears symmetrically. The saddle-point solution in the case
of the complex curvature matrix model is identical in most respects to that of
the Hermitian models above. Now the spectral density $\rho_C(h)$ for the Young
tableau weights $h_i^{(0)}$, $i=1,\dots,\frac N3$, obeys the saddle-point
equation
\beq
\pvint_{\!\!b}^a~dh'~\frac{\rho_C(h')}{h-h'}=-\log(\lambda^{3/2}h)-
\log\left(\frac{h}{h-b}\right)-\frac{1}{2}{\cal F}(h)~~~~~,~~~~~h\in[b,a]
\label{saddlepteqC}\eeq
where the logarithmic derivative ${\cal F}(h)$ of the Itzykson-Zuber integral
is defined just as in (\ref{izcorr}). The solution for ${\cal F}(h)=0$ is thus
identical to those above. In particular, working out the free energy as before
we see that
\beq
S_C[\lambda,\tilde t_q^*]=4S_H[\rho_C;\sqrt{\lambda},\tilde
t_q^*]-\int_0^adh~\rho_C(h)\log I_c[h,\bar A]
\label{freeCHrel}\eeq

The combinatorical factors appearing in (\ref{freeCHrel}) can be understood
from the Wick expansions of the Hermitian and complex curvature matrix models.
First consider the case ${\cal F}(h)=0$. One factor of 2 appears in front of
$S_H$ in (\ref{freeCHrel}) because the number of independent degrees of freedom
of the $N\times N$ complex matrix model is twice that of the $N\times N$
Hermitian curvature matrix model. The other factor of 2 arises from the mapping
of Feynman graphs of the complex matrix model onto those of the Hermitian model
(Fig. 4). At each 6-point vertex of a complex graph we can place a
$\phi^\dagger$ line beside a $\phi$ line to give a graph with 3-point vertices
and ``thick" lines (each thick line associated with a $\phi^\dagger\phi$ pair
of lines). This maps the propagator weights as $\lambda^6\to\lambda^3$ and
there are $3!/3=2$ distinct ways of producing a complex graph by thickening and
splitting the lines of a graph in ${\cal M}_3$ in this way. This is the
relation (\ref{freeCHrel}) for $I_c\equiv1$. In the general case, there is a
relative factor of 4 in front of the Itzykson-Zuber correlator in
(\ref{saddlepteqC}) because the Feynman rules for the complex curvature matrix
model associate the weights
\beq
t_q^{C*}=\frac1q\frac\tr N\left(A^{1/2}\right)^q
\label{complexweights}\eeq
to the dual vertices $v_q^*$ of the Feynman graphs (compare with
(\ref{dualweights})). Thus for $q\tilde t_q^*\neq1$, the free energy of the
complex model differs from (\ref{SHsaddlept}) in a factor of 4 in front of the
integral involving the Itzykson-Zuber integral $I_c[h,\bar A]$.

The complex curvature matrix model with the above rules thus generates
``checkered" Riemann surfaces corresponding to the double-lined triangulations
shown in Fig. 4. A 3-colouring of a graph of ${\cal M}_3$ is now described by a
sort of 6-colouring whereby each colour R, B, and G is assigned an incoming and
outgoing orientation at each vertex. Again these relations are only valid at
$N=\infty$. However, the models lie in the same universality class (as with the
even coordination number models discussed above) and consequently their double
scaling limits will be the same. In particular, the equivalences
(\ref{freerels}) and (\ref{freeCHrel}) will hold to all genera near the
critical point \cite{ackm} (see the next subsection).

\begin{figure}
\unitlength=0.90mm
\linethickness{0.4pt}
\begin{center}
\begin{picture}(100.00,35.00)(0,10)
\small
\thicklines
\put(35.00,25.00){\circle*{2.50}}
\put(80.00,25.00){\circle*{2.50}}
\put(25.00,35.00){\line(1,-1){22}}
\put(25.00,15.00){\line(1,1){22}}
\put(21.00,25.00){\line(1,0){28}}
\put(24.00,24.50){\makebox(0,0){\Large$\to$}}
\put(42.00,24.50){\makebox(0,0){\Large$\to$}}
\put(25.00,35.00){\makebox(0,0){\Large$\searrow$}}
\put(25.00,15.00){\makebox(0,0){\Large$\nearrow$}}
\put(40.00,30.00){\makebox(0,0){\Large$\nearrow$}}
\put(40.00,20.00){\makebox(0,0){\Large$\searrow$}}
\put(60.00,25.00){\makebox(0,0){$\longrightarrow$}}
\put(81.00,25.00){\line(0,1){15}}
\put(79.00,25.00){\line(0,1){14}}
\put(79.00,25.00){\line(-1,-1){10}}
\put(81.00,25.00){\line(-1,-1){12}}
\put(79.00,25.00){\line(1,-1){10}}
\put(81.00,25.00){\line(1,-1){12}}
\put(79.00,37.00){\makebox(0,0){\Large$\downarrow$}}
\put(81.00,32.00){\makebox(0,0){\Large$\uparrow$}}
\put(70.00,16.00){\makebox(0,0){\Large$\nearrow$}}
\put(76.00,20.00){\makebox(0,0){\Large$\swarrow$}}
\put(89.00,15.00){\makebox(0,0){\Large$\nwarrow$}}
\put(86.00,20.00){\makebox(0,0){\Large$\searrow$}}
\end{picture}
\end{center}
\begin{description}
\small
\baselineskip=12pt
\item[Figure 4:] The mapping of a complex model 6-point vertex onto a Hermitian
model 3-point vertex. The incoming lines represent the $\phi^\dagger$ fields,
the outgoing lines the $\phi$ fields, and the solid circle denotes an insertion
of the matrix $A^{1/2}$.
\end{description}
\end{figure}

\subsection{Critical Behaviour and Correlation Functions}

We will now discuss to what extent the above models are universal. First, let
us find the explicit solution of the saddle-point equation (\ref{saddlepteq})
in the case ${\cal F}(h)=0$ \cite{kaz1} (i.e. $\bar A=\tilde A={\bf1}$ and
$q\tilde t_q^*=1$). Notice that in this case the dual potentials in
(\ref{VevenXA}) and (\ref{V3*}) become the Penner-type potentials
\cite{chekmak,di}
\beq\new{\begin{array}{c}
V_{\rm even}(XA)=-\frac{1}{2}\log\left({\bf1}-(XA^{(2)})^2\right)~~~~~,~~~~~
V_4^{(*)}(XA)=-\frac12\log\left({\bf1}-(XA_4)^2\right)\\V_3^{(*)}(XA)=-\frac13
\log\left({\bf1}-(XA_3)^3\right)\end{array}}
\label{Pennerpots}\eeq
so that the above arguments imply the equivalences between the curvature matrix
model for a dynamical triangulation and those of Penner models. The behaviour
of the saddle-point solution when the dual vertices $v_q^*$ are not weighted
equally will be examined in the next section.

The saddle-point equation (\ref{saddlepteq}) determines the continuous part of
the resolvent function ${\cal H}_H(h)$ across its cut. Using the normalization
$\int_0^adh~\rho_H(h)=1$ of the spectral density in (\ref{resolv}) implies the
asymptotic boundary condition ${\cal H}_H(h)\sim1/h$ at $|h|\to\infty$. It
follows that the solution for the resolvent function is given by \cite{fgz}
\beq\new{\begin{array}{lll}
{\cal H}_H(h)&=&\log\left(\frac{h}{h-b}\right)-\oint_{\cal C}\frac{ds}{2\pi
i}~\frac{1}{s-h}\sqrt{\frac{(h-a)(h-b)}{(s-a)(s-b)}}~\pvint_{\!\!b}^adh'~
\frac{\rho_H(h')}{s-h'}\\&=&
\log\left(\frac{h}{h-b}\right)+\oint_{\cal C}\frac{ds}{2\pi
i}~\frac{1}{s-h}\sqrt{\frac{(h-a)(h-b)}{(s-a)(s-b)}}\left\{\log(\lambda^3s)+
\log\left(\frac{s}{s-b}\right)\right\}\end{array}}
\label{ressoln}\eeq
The contour integrations in (\ref{ressoln}) can be evaluated by blowing up the
contour $\cal C$ to infinity and catching the contributions from the cuts of
the two logarithms, on $(-\infty,0]$ and $[0,b]$, respectively, and also from
the simple pole at $s=h$. Keeping careful track of the signs of the square
roots on each cut and taking discontinuities across these cuts, after some
algebra we arrive at
\beq\new{\begin{array}{lll}
{\cal H}_H(h)&=&-\log(\lambda^3h)+\sqrt{(h-a)(h-b)}\left\{\int_0^b
-\int_{-\infty}^0\right\}\frac{dx}{x-h}~\frac{1}
{\sqrt{(x-a)(x-b)}}\\&=&\log\left\{\frac{a-b}{\lambda^3(\sqrt a+\sqrt
b)^2}\frac{\left(h+\sqrt{ab}+\sqrt{(h-a)(h-b)}\right)^2}{h\left(
h(a+b)-2ab+2\sqrt{ab(h-a)(h-b)}\right)}\right\}\end{array}}
\label{resexactHt=1}\eeq

It remains to solve for the endpoints $a,b$ of the support of the spectral
distribution function in terms of the parameters of the matrix potential. They
are determined by expanding the function (\ref{resexactHt=1}) for large-$h$,
and requiring that the constant term vanish and that the coefficient of the
$1/h$ term be 1. This leads to a pair of equations for $a,b$
\beq
\xi=\frac13(\eta-1)~~~~~,~~~~~3\lambda^6\eta^3=\eta-1
\label{bdryeq1}\eeq
where we have introduced the positive endpoint parameters
\beq
\xi=\frac14\left(\sqrt a-\sqrt b\right)^2~~~~~,~~~~~\eta=\frac14\left(\sqrt
a+\sqrt b\right)^2
\label{xietadef}\eeq
The cubic equation for $\eta$ in (\ref{bdryeq1}) can be solved explicitly. The
original matrix model free energy is analytic around $\lambda=0$. We shall see
below that it is analytic in $\eta$, so that we should choose the branch of the
cubic equation in (\ref{bdryeq1}) which is regular at $\lambda=0$. This
solution is found to be
\beq
\eta=-\frac12\beta^{1/3}-\frac1{18\lambda^6\beta^{1/3}}-i\sqrt3
\left(\beta^{1/3}-\frac1{9\lambda^6\beta^{1/3}}\right)
\label{etaregsol}\eeq
where
\beq
\beta=-\frac1{6\lambda^6}+\frac1{54\lambda^9}\sqrt{81\lambda^6-4}
\label{betadef}\eeq
The solution (\ref{etaregsol}) leads to the appropriate boundary condition
$\eta(\lambda=0)=1,\xi(\lambda=0)=0$, i.e. $a=b=1$, as expected since in this
case the spectral density should be trivial, $\rho_H(h)\equiv1$.

The solution (\ref{etaregsol}) is real-valued for
$\lambda<\lambda_c\equiv(2/9)^{1/3}$ which is the regime wherein all three
roots of the cubic equation for $\eta$ in (\ref{bdryeq1}) are real. For
$\lambda>\lambda_c$ there is a unique real-valued solution which does not
connect continuously with the real solution that is regular at $\lambda=0$. At
$\lambda=\lambda_c$ the pair of complex-conjugate roots for $\lambda>\lambda_c$
coalesce and become real-valued. Thus for $\lambda>\lambda_c$ the one-cut
solution above for ${\cal H}_H$ is no longer valid and the spectral density
$\rho_H$ becomes supported on more than one interval. The value
$\lambda=\lambda_c$ is therefore identified as the critical point of a phase
transition of the random surface theory which corresponds to a change of
analytic structure of the saddle-point solution of the matrix model. As usual,
at this critical point the continuum limit of the discretized random surface
model, describing the physical string theory, is reached. To determine the
precise nature of the geometry that is obtained in this continuum limit, we
need to study the scaling behaviour of the matrix model free energy near the
critical point. In the following we shall present two calculations illustrating
this critical behaviour.

\subsubsection{Propagator}

As shown in \cite{kaz1,kaz1a}, in the general case the resolvent function
${\cal H}_H(h)$ and the Itzykson-Zuber correlator ${\cal F}(h)$ completely
determine the solution of the random matrix model in the large-$N$ limit. For
instance, expectation values of the operators $\frac\tr NX^{2q}$ can be
obtained from the contour integral
\beq
\left\langle\frac\tr N X^{2q}\right\rangle_{{\cal
M}_3}=\frac{\lambda^q}q\oint_{\cal C}\frac{dh}{2\pi i}~h^q\e^{q{\cal H}_H(h)}
\label{trNX2qH}\eeq
where the normalized average is with respect to the partition function
(\ref{partfn}) and the decomposition (\ref{A3}). The expression (\ref{trNX2qH})
holds only for the even moments of the matrix $X$ because it is only for such
moments that the character expansion of the right-hand side of (\ref{trNX2qH})
admits a symmetrical even-odd decomposition as in the Itzykson-Di Francesco
formula \cite{kaz1}. Although a character expansion for the odd moments can be
similarly derived, the weights will not split symmetrically and there does not
appear to be a general expression for them in terms of the weight distribution
function at large-$N$. For the even coordination number models above, the
reflection symmetry $X\to-X$ of the matrix model forces the vanishing of all
odd moments automatically and (\ref{trNX2qH}) represents the general expression
for the complete set of observables which do not involve the external field $A$
at large-$N$. In terms of the character expansion formula, the character sum
for $\langle\frac\tr NX^{2q+1}\rangle$ vanishes in the case of an even
potential because of the localization onto the even representations of
$GL(N,{\Bbb C})$, whereas in the 3-point model $\langle\frac\tr
NX^{2q+1}\rangle_{{\cal M}_3}\neq0$ because its character sum is supported on
mod 3 congruence class representations.

The contour integral representation (\ref{trNX2qH}), with ${\cal H}_H$ replaced
by ${\cal H}_C$, also holds for the correlators $\langle\frac\tr
N(\phi^\dagger\phi)^q\rangle_C$ of the complex curvature matrix model. These
averages form the complete set of $A$-independent observables at large-$N$ in
this case because of the charge conjugation symmetry $\phi\to\phi^\dagger$ of
the complex model (which is the analogue here of the reflection symmetry of the
even coordination number models). Again this indicates that the complex
curvature matrix model is better suited to represent the triangulated random
surface sum, because it is more amenable to an explicit solution in the
large-$N$ limit. Furthermore, its complete set of correlation functions
coincide at $N=\infty$ with those of the equivalent matrix models defined above
with even potentials, after redefining the vertex weights according to
(\ref{dualweights}) and (\ref{complexweights}).

The contour integral on the right-hand side of (\ref{trNX2qH}) can be evaluated
by blowing up the contour $\cal C$, expanding the resolvent function
(\ref{resexactHt=1}) for large-$h$, and computing the residue at $h=\infty$.
For example, for $q=1$ we have
\beq
\left\langle\frac\tr NX^2\right\rangle_{{\cal
M}_3}=\lambda\left(\frac12+\langle h\rangle\right)
\label{trNX2H}\eeq
where the weight average $\langle
h\rangle=\langle\frac3N\sum_{i=1}^{N/3}h_i^{(0)}\rangle=\int_0^adh~\rho_H(h)h$
corresponds to the coefficient of the $1/h^2$ term in the asymptotic expansion
of (\ref{resexactHt=1}). This result also follows directly from differentiating
the free energies
\beq
\left\langle\frac\tr
NX^2\right\rangle=2\lambda^2\frac{\partial}{\partial\lambda}S[\lambda,t_q^*]
{}~~~~~,~~~~~\left\langle\frac\tr
N\phi^\dagger\phi\right\rangle_C=\lambda^2\frac{\partial}{\partial\lambda}
S_C[\lambda,t_q^*]
\label{propsfromfree}\eeq
and using the saddle-point equations. Notice that (\ref{propsfromfree}) and the
free energy relations (\ref{freerels}) and (\ref{freeCHrel}) imply, in
particular, that the matrix propagators of the 3-point, 4-point, even-even and
complex matrix models coincide with the appropriate redefinitions of the
coupling constants $\lambda$ and $t_q^*$. However, the equivalences of generic
observables in the various models will not necessarily hold. We shall return to
this point shortly.

Using (\ref{resexactHt=1}) and (\ref{bdryeq1}), the propagator (\ref{trNX2H})
can be written as
\beq
\left\langle\frac\tr NX^2\right\rangle_{{\cal
M}_3}=\lambda\left(\frac13-\frac{\eta^2}3+\eta\right)
\label{trNX2Hexpleta}\eeq
Expanding (\ref{trNX2Hexpleta}) using (\ref{etaregsol}) as a power series in
$\lambda$ yields
\beq
\left\langle\frac\tr NX^2\right\rangle_{{\cal
M}_3}=\lambda+\lambda^7+6\lambda^{13}+54\lambda^{19}+594\lambda^{25}+{\cal
O}(\lambda^{31})
\label{propexp}\eeq
We have verified, by an explicit Wick expansion of the matrix propagator (up to
and including the order shown in (\ref{propexp})) using the $3m$-sided polygon
constraints on the matrix traces (\ref{dualweights}), that (\ref{propexp})
indeed coincides with perturbative expansion of the curvature matrix model
(\ref{partfn}) and thus correctly counts the planar 3-point fat graphs
consisting of only $3m$-sided polygons. It also agrees with the Wick expansions
of the even coordination number models above \cite{kaz1} and of the complex
curvature matrix model.

To examine the scaling behaviour of observables of the matrix model near the
critical point, we introduce a renormalized, continuum cosmological constant
$\Lambda$ and momentum $\Pi$ by
\beq
\lambda^6=\lambda_c^6(1-\Lambda)~~~~~~,~~~~~~\eta=\eta_c-\Pi/2
\label{contdefs}\eeq
where $\eta_c=\eta(\lambda=\lambda_c)=3/2$. We then approach the critical point
along the line $\Lambda,\Pi(\Lambda)\to0$, where the function $\Pi(\Lambda)$ is
found by substituting the definitions (\ref{contdefs}) into (\ref{etaregsol})
to get
\beq
\Pi(\Lambda)=-\frac12\left[\Xi^{1/3}+\frac{9(\Lambda_*+1)}{\Xi^{1/3}}-6+
i\sqrt3\left(\Xi^{1/3}-\frac{9(\Lambda_*+1)}{\Xi^{1/3}}\right)\right]
\label{PiLambda}\eeq
with
\beq
\Xi=27\left(\Lambda_*+1+\sqrt{-\Lambda_*^3-2\Lambda_*^2-\Lambda_*}\right)
{}~~~~~,~~~~~\Lambda_*=\frac{\Lambda}{1-\Lambda}
\label{Xi*def}\eeq
Substituting (\ref{contdefs}) and (\ref{PiLambda}) into (\ref{trNX2Hexpleta})
and expanding around $\Lambda=0$ yields after some algebra
\beq
\left\langle\frac\tr NX^2\right\rangle_{{\cal
M}_3}=\frac29\Lambda^{3/2}+\dots\equiv c\cdot\Lambda^{1-\gamma_{\rm str}}+\dots
\label{corrcrit}\eeq
where the dots denote terms which are less singular as $\Lambda\to0$. The
leading non-analytic behaviour in (\ref{corrcrit}) identifies the critical
string exponent of this random surface model as
\beq
\gamma_{\rm str}=-1/2
\label{gravstr}\eeq
so that the system in the continuum limit represents pure two-dimensional
quantum gravity \cite{fgz}. The same argument also applies to the 4-point and
even-even models with the appropriate redefinitions of cosmological constants
in (\ref{contdefs}). Thus the equal weighting of all vertices in these random
surface models leads to curvature matrix models in the same universality class
as the more conventional Hermitian one-matrix models of two-dimensional quantum
gravity. This agrees with recent numerical results in \cite{bct}.

\subsubsection{General Correlators}

We now turn to a more general discussion of the evaluation of observables in
the matrix models above. The relations between correlators involving the
external matrix $A$ are a bit more subtle than those represented by the contour
integrations in (\ref{trNX2qH}) which coincide in all four matrix models for
all $q$. Because of the simple line maps that exist between the ensembles
${\cal M}_3^*$, ${\cal M}_4^*$ and ${\cal M}_{\rm even}$, all expectation
values are the same in these models, i.e.
\beq
\frac14\left\langle\frac\tr N(\phi^\dagger\phi
A_3)^{3q}\right\rangle_{C^*}=\left\langle\frac\tr
N(XA_3)^{3q}\right\rangle_{{\cal M}_3^*}
=\left\langle\frac\tr N(XA_4)^{2q}\right\rangle_{{\cal M}_4^*}
=\left\langle\frac\tr N(XA^{(2)})^{2q}\right\rangle_{{\cal M}_{\rm even}}
\label{even4*corrrel}\eeq
Analytically, this equality follows from the fact that these correlators are
given by derivatives of the free energies of the matrix models with respect to
the weights $\tilde t_q^*$. The powers in (\ref{even4*corrrel}) can be
understood from the fact that the reflection symmetry $A^{(2)}\to-A^{(2)}$ of
the ${\cal M}_4$ and ${\cal M}_{\rm even}$ models (which restrict the non-zero
averages to $\langle\frac\tr N(XA^{(2)})^{2q}\rangle_{{\cal M}_4,{\cal M}_{\rm
even}}$) corresponds to the ${\Bbb Z}_3$-symmetry $A^{(3)}\to\omega_3A^{(3)}$,
$\omega_3\in{\Bbb Z}_3$, of the ${\cal M}_3$ model (which restricts the
non-vanishing observables to $\langle\frac\tr N(XA^{(3)})^{3q}\rangle_{{\cal
M}_3}$ and $\langle\frac\tr N(\phi^\dagger\phi A^{(3)})^{3q}\rangle_C$). In
terms of the coloured graphs of subsection 3.2, these discrete symmetries
correspond to permutations of the colours of graphs in each ensemble. It is
these symmetries that are ultimately responsible for the localization of the
Itzykson-Di Francesco character sum onto the appropriate even or mod 3
representations in each case.

The correlators in (\ref{even4*corrrel}) are also given by a simple contour
integration. It can be shown that \cite{kaz1a}
\beq
\frac1N\frac\partial{\partial\tilde
t_q^*}\log\chi_{\left\{\frac{h^{(0)}}3\right\}}(\bar
A)=\sum_{k=1}^{N/3}\frac{\chi_{\left\{\tilde h^{(0)}_k(-q)/3\right\}}(\bar
A)}{\chi_{\left\{h^{(0)}/3\right\}}(\bar A)}
\label{charderivsum}\eeq
where $\tilde h_k^{(0)}$ is defined in (\ref{tildehk}). From the character
expansion of section 2 we see that the left-hand side of (\ref{charderivsum})
arises from a derivative of the dual model free energy
$S_H^{(*)}[\lambda,\tilde t_q^*]$ with respect to the vertex weights $\tilde
t_q^*$, whereas the right-hand side is identical to (\ref{weightNid}) with
$q\to-q$. In the large-$N$ limit, we can therefore represent the dual model
correlators in (\ref{even4*corrrel}) by the contour integrations
\beq
\left\langle\frac\tr N(XA_3)^{3q}\right\rangle_{{\cal
M}_3^*}=-\frac{\lambda^q}q\oint_{\cal C}\frac{dh}{2\pi i}~\e^{-q({\cal
H}_H(h)+{\cal F}(h))}
\label{3ptdualcorrs}\eeq
Although it is possible to determine the complete set of observables of the
dual matrix models in terms of the saddle-point solution at large-$N$, the
situation is somewhat more complicated for the correlators of the ${\cal M}_3$
and ${\cal M}_4$ ensembles. The above discussion illustrates to what extent the
large-$N$ saddle-point solution of the Itzykson-Di Francesco formula can be
used to represent the observables of the matrix model. Those which do admit
such a representation typically appear to be obtainable from derivatives of the
free energy of the model and thus correspond in the random surface
interpretation to insertions of marked loops on the surfaces. Thus, strictly
speaking, the natural observable to compute in the large-$N$ limit of the
3-point matrix model is the free energy (\ref{SHsaddlept}). In the next
subsection we shall show how this calculation carries through.

\subsubsection{Free Energy}

The natural observable to compute in the triangulation model is the large-$N$
(genus zero) free energy (\ref{SHsaddlept}) with $I_c=1$. Splitting up the
integration range and setting $\rho_H(h)=1$ on $[0,b]$, we find
\beq\new{\begin{array}{c}
S_H=\frac13\int_b^adh
{}~\rho_H(h)h\left[\log(\lambda^3h)-1\right]+\frac13\int_b^adh~\rho_H(h)
\left[h\log h-(h-b)\log(h-b)-b\right]\\
+\frac16\int_b^a\!\pvint^a_{\!\!b}dh~dh'~\rho_H(h)
\rho_H(h')\log|h-h'|+\frac{b^2}6\left(\log
b-\frac32\right)+\frac{b^2-3/2}4\log\lambda\end{array}}
\label{freeints}\eeq
The spectral density is found by computing the discontinuity (\ref{disceq}) of
the weight resolvent function (\ref{resexactHt=1}) across the cut $[b,a]$,
which yields
\beq
\rho_H(h)=\frac1\pi\left[\arctan\left(\frac{2\sqrt{ab(a-h)(h-b)}}{(a+b)h-2ab}
\right)-2\arctan\left(\frac{\sqrt{(a-h)(h-b)}}{h+\sqrt{ab}}\right)\right]
{}~~~~,~~~h\in[b,a]
\label{rhoHexpl}\eeq

The double integral in the free energy (\ref{freeints}) can be simplified by
integrating up the saddle-point equation (\ref{saddlepteq}) for $h\in[b,a]$ to
get
\beq\new{\begin{array}{lll}
\pvint_{\!\!b}^adh'~\rho_H(h')\log|h-h'|&=&h\left[1-\log(\lambda^3
h)\right]+(h-b)\log(h-b)-h\log h+\frac14\log\lambda\\&
&+\pvint_{\!\!b}^adh'~\rho_H(h')\log(h'-b)+b\left[\log(\lambda^3b)-1\right]
+b\log b\end{array}}
\label{saddlepteqint}\eeq
Substituting (\ref{saddlepteqint}) into (\ref{freeints}) and integrating by
parts, we find after some algebra
\beq\new{\begin{array}{c}
S_H=-\frac16\int_b^adh~\frac{d\rho_H(h)}{dh}\left\{h\left[h\log
h-\frac{h}2\log(h-b)+\log(h-b)\right]-b\left(1-\frac
b2\right)\log(h-b)\right\}\\+\frac b3\log b\left(1-\frac
b2\right)+\frac{b}6\log\lambda-\frac b{12}\left(b+15\right)+\frac{\langle\tilde
h\rangle}6\left(\log\lambda-\frac32\right)\end{array}}
\label{SPdrho}\eeq
where we have introduced the reduced weight average $\langle\tilde
h\rangle=\int_b^adh~\rho_H(h)h=\langle h\rangle-b^2/2$, and from
(\ref{rhoHexpl}) we have
\beq
\frac{d\rho_H(h)}{dh}=\frac{h-2\sqrt{ab}}{\pi\sqrt{(a-h)(h-b)}}~~~~~,~~~~~
h\in[b,a]
\label{drhoPdh}\eeq
The weight average $\langle h\rangle$ can be read off from (\ref{resexactHt=1})
to give
\beq
\langle h\rangle=-\eta^2/3+\eta-1/6
\label{havgpen}\eeq

To evaluate the remaining logarithmic integrals in (\ref{SPdrho}), it is
convenient to change variables from $h\in[b,a]$ to $x\in[-1,1]$ with
$h=\frac12(a+b)+\frac12(a-b)x$. This leads to a series of integrals over
elementary algebraic forms and forms involving logarithmic functions. The
latter integrals can be computed by rewriting the integrations over
$x\in[-1,1]$ as contour integrals, blowing up the contours and then picking up
the contributions from the discontinuities across the cuts of the logarithms.
The relevant integrals are then found to be
\beq\new{\begin{array}{l}
I_0(r)=\int_{-1}^1dx~\frac{\log(1+rx)}{\sqrt{1-x^2}}=\pi\log\left(
\frac{1+\sqrt{1-r^2}}2\right)\\I_1(r)=\int_{-1}^1dx~\frac{x\log(1+rx)}
{\sqrt{1-x^2}}=\frac\pi
r\left(1-\sqrt{1-r^2}\right)\\I_2(r)=\int_{-1}^1dx~\frac{x^2\log(1+rx)}
{\sqrt{1-x^2}}=\frac12\left(I_0(r)-\frac{I_1(r)}r+\frac\pi2\right)\\J(r)=
\int_{-1}^1dx~\frac{\log(1+x)}{(1+rx)\sqrt{1-x^2}}
=-\pi\pvint_{\!\!-\infty}^{-1}
\frac{dy}{(1+ry)\sqrt{1-y^2}}+\frac{\pi
\log\left(\frac{1-r}r\right)}{\sqrt{1-r^2}}
\\~~~~~~~~~~~~~~~~~~~~~~~~~~~~~~~~~~~~~~~~~~
=\frac\pi{\sqrt{1-r^2}}\log\left(\frac{1-r}{1+\sqrt{1-r^2}}\right)\end{array}}
\label{logintids}\eeq
where $0\leq r\leq1$. Using the above identities and the boundary conditions
(\ref{bdryeq1}), after some tedious algebra we arrive finally at the simple
expression
\beq
S_H(\eta)=\frac14\log\lambda+\frac16\log\eta+\frac{\eta^2}{36}-\frac{7\eta}{36}
\label{SPetafinal}\eeq
for the free energy of the triangulation model, where we have ignored an
overall numerical constant.

To examine the scaling behaviour of the free energy about the critical point,
we once again introduce a renormalized, continuum cosmological constant
$\Lambda$ and momentum $\Pi$ as in (\ref{contdefs}). From the boundary equation
$3\lambda^6\eta^3=\eta-1$ we can calculate derivatives of $\eta$ with respect
to the cosmological constant to get
\beq
\frac{\partial\eta}{\partial\Lambda}=-\frac{3\lambda_c^6\eta^4}{\eta-\eta_c}
\label{deriveta}\eeq
which we note diverges at the critical point. From (\ref{SPetafinal}) and
(\ref{deriveta}) the first two derivatives of the free energy are then found to
be
\beq
\frac{\partial S_H}{\partial\Lambda}=-\lambda_c^6\eta^3(\eta-2)/6
{}~~~~~,~~~~~\frac{\partial^2S_H}{\partial\Lambda^2}=2\lambda_c^{12}\eta^6
\label{freederivsL}\eeq
Both derivatives in (\ref{freederivsL}) are finite at the critical point.
Taking one more derivative yields a combination that does not cancel the
singular $(\eta-\eta_c)^{-1}$ part of (\ref{deriveta}), i.e. there is a third
order phase transition at the critical point. Substituting (\ref{contdefs})
into the boundary equation (\ref{bdryeq1}) gives $\Pi^2\sim\Lambda$ near
criticality $\Lambda\to0$ (see (\ref{PiLambda}),(\ref{Xi*def})), and so
substituting $\eta\sim\eta_c-\Lambda^{1/2}$ into (\ref{freederivsL}) and
expanding about $\Lambda=0$ yields
\beq
\frac{\partial^2S_H}{\partial\Lambda^2}=c\cdot\Lambda^{1/2}+\dots
\label{SHexpcrit}\eeq
which identifies the expected pure gravity string exponent (\ref{gravstr}).

\section{Incorporation of the Itzykson-Zuber Correlator}

In this section we shall give a more quantitative presentation of the
evaluation of observables in the curvature matrix models and how they reproduce
features of the triangulated random surface sum. This will also further
demonstrate the validity of the weight splitting that was assumed to hold in
the Itzykson-Di Francesco formula, i.e. that the leading order contributions to
the partition function indeed do localize at $N=\infty$ onto the even
representations of $GL(N,{\Bbb C})$ that split symmetrically into mod 3
congruence classes. For this we shall present an explicit evaluation of the
Itzykson-Zuber correlator ${\cal F}(h)$ using (\ref{weightcont}). This will
amount to an evaluation of the large-$N$ limit of the generalized Schur
functions introduced in the previous section.

Notice first that expanding both sides of (\ref{weightcont}) in powers of $q$
and equating the constant terms leads to the identity
\beq
1=\oint_{\cal C}\frac{dh}{2\pi i}~\{{\cal H}_H(h)+{\cal F}(h)\}
\label{aneq}\eeq
which, along with the normalization $\int_0^adh~\rho_H(h)=1$ of the spectral
density, implies that
\beq
\oint_{\cal C}dh~{\cal F}(h)=0
\label{Fhanalytic}\eeq
Thus, at least for some range of the couplings $\tilde t_q^*$, the function
${\cal F}(h)$ will be analytic in a neighbourhood of the cut of ${\cal
H}_H(h)$. This will be true so long as the Itzykson-Zuber integral does not
undergo a phase transition which changes its analyticity features in the
large-$N$ limit. The equation (\ref{weightcont}) can be used to determine
${\cal F}(h)$ once the coupling constants of the dynamical triangulation are
specified. The simplest choice is a power law variation of the couplings as the
coordination numbers are varied,
\beq
q\tilde t_q^*=t^{q-1+p}~~~~~~;~~~~~~t\in{\Bbb R}~,~p\in{\Bbb Z}
\label{weightchoice}\eeq
so that each vertex of the triangulated surface interacts with a strength
proportional to $t$ (with proportionality constant $t^p$) with each of its
nearest neighbours.

This simple choice of vertex weights in the triangulation model allows us to
study more precisely how the curvature matrix model represents features of the
random surface sum, and to further demonstrate that the actual saddle-point
localization of the partition function is not onto some configuration of Young
tableau weights other than the mod 3 representations. It will also allow us to
examine how the observables of the model behave as the weights are varied in
this simple case, without the complexities that would appear in the analytic
form of the solution for other choices of coupling constants. For generic
$t\neq1$ (i.e. ${\cal F}(h)\neq0$), the saddle-point solution of the matrix
model must satisfy certain physical consistency conditions so that it really
does represent the (continuum) genus zero contribution to the random surface
sum (\ref{partgraphs}) with the choice of couplings (\ref{weightchoice}). Using
Euler's theorem $V-E+F=2$ for a spherical topology and the triangulation
relation
\beq
\sum_{v_q^*\in G_3}q=2E=3F
\label{triangrel}\eeq
it follows that the planar surface sum in (\ref{partgraphs}) is
\beq
Z^{(0)}_H(\lambda,t)=t^{2(p-1)}\sum_{G_3^{(0)}}c_{G_3^{(0)}}\left(
\lambda^3t^{p+1}\right)^{A(G_3^{(0)})}
\label{partsumexpl}\eeq
where the sum is over all planar fat-graphs $G_3^{(0)}$ of area
$A(G_3^{(0)})\propto F(G_3^{(0)})$, and the constant $c_{G_3^{(0)}}$ is
independent of the coupling constants $\lambda$ and $t$. The perturbative
expansion parameter is $\lambda^6t^{2(p+1)}$ and the critical point, where a
phase transition representing the continuum limit of the discretized random
surface model takes place, is reached by tuning the expansion parameter to the
radius of convergence of the power series (\ref{partsumexpl}). The critical
line $\lambda_c(t)$ thus obeys an equation of the form
\beq
\lambda_c(t)^6\cdot t^{2(p+1)}=~{\rm constant}
\label{critline}\eeq
The solution of the model for $t\neq1$ should therefore have the same physical
characteristics as that with $t=1$, since in the random surface interpretation
of the curvature matrix model the only effect of changing $t$ is to rescale the
cosmological constant of the random surface sum for $t=1$ as
$\lambda^3\to\lambda^3t^{p+1}$. This geometrical property should be reflected
in the large-$N$ solution of the matrix model with a non-vanishing
Itzykson-Zuber correlator.

As discussed in \cite{kaz1a}, it is possible to invert the equation
\beq
G(h)=\e^{{\cal H}_H(h)+{\cal F}(h)}
\label{Gdef}\eeq
to obtain $h$ as a function of $G$, by changing variables in the contour
integral (\ref{weightcont}) from $h$ to $G$ to get
\beq
q\tilde t_q^*=\oint_{{\cal C}_G}\frac{dG}{2\pi iG}~h(G)G^q
\label{weightinv}\eeq
where the contour ${\cal C}_G$ encircles the cut $[G(b),0]$ with clockwise
orientation in the complex $G$-plane. Then
\beq
h(G)=1+\sum_{q=1}^\infty\frac{q\tilde t_q^*}{G^q}+\sum_{q=1}^\infty g_qG^q
\label{hG}\eeq
where the coefficients
\beq
g_q\equiv\oint_{{\cal C}_G}\frac{dG}{2\pi
iG}~h(G)G^{-q}=\lambda^{-q}\left\langle\frac\tr
N(XA_3)^{3q}\right\rangle_{{\cal M}_3^*}
\label{gq}\eeq
determine the analytic part of the function $h(G)$. The second equality in
(\ref{gq}) follows from (\ref{3ptdualcorrs}) and the constant term in the
Laurent series expansion (\ref{hG}) is unity because of the normalization of
the spectral density $\rho_H$ \cite{kaz1a}. In the general case, the solution
for $G(h)$ as determined from (\ref{hG}) will be multi-valued. It was shown in
\cite{kaz1a} that the first set of sheets of the Riemann surface of this
multi-valued function are connected along, and hence determined by, the cut
structures of $\e^{{\cal F}(h)}$, which map the point $h=\infty$ to $G=0$.
These sheets are found by inverting (\ref{hG}) with $g_q=0$. The remaining
sheets are determined by the cuts of $\e^{{\cal H}_H(h)}$ and are associated
with the positive powers of $G$ in (\ref{hG}).

The solution is extremely simple, however, for the choice of couplings
(\ref{weightchoice}), as then the inversion of the equation (\ref{hG}) with
$g_q=0$ yields
\beq
G_1(h)=t+\frac{t^p}{h-1}
\label{sheet1}\eeq
This solution has a simple pole of residue $t^p$ at $h=1$, but no multivalued
branch cut structure. Thus we expect that the singularities of $\e^{{\cal
F}(h)}$ will have a pole structure, rather than a cut structure. The remaining
sheets of the function $G(h)$ will be determined by the cut structure of
$\e^{{\cal H}_H(h)}$. They are attached to the ``physical" sheet $G_1(h)$, on
which the poles of $\e^{{\cal F}(h)}$ and the cuts of $\e^{{\cal H}_H(h)}$ lie,
by these cuts. Note that the analytical structure of the character as
determined by this part of the Laurent series in (\ref{hG}) is anticipated from
the Schur character formula (\ref{schurchar}),(\ref{schurpoly}).

With this observation we can in fact obtain a closed form expression for the
Itzykson-Zuber correlator in the large-$N$ limit, in contrast to the generic
case \cite{kaz1a} where in general one obtains only another discontinuity
equation such as (\ref{saddlepteq}). The resolvent function (\ref{resolv}) can
be written as
\beq
{\cal H}_H(h)=\log\left(\frac{h}{h-b}\right)+\tilde{\cal H}_H(h)
\label{Hsplit}\eeq
where $\tilde{\cal H}_H(h)=\int_b^adh'~\rho_H(h')/(h-h')$ is the reduced
resolvent function associated with the non-trivial part of the density
$\rho_H$, and it has a branch cut on the interval $[b,a]$. Using
(\ref{Fhanalytic}) it can be written as
\beq
\tilde{\cal H}_H(h)=\oint_{\tilde{\cal C}}\frac{d\tilde h}{2\pi i}~\frac{\log
G(\tilde h)}{h-\tilde h}=-\oint_{{\cal C}_G}\frac{dG}{2\pi
i}~\frac{h'(G)}{h-h(G)}\log G
\label{resnontr}\eeq
where the contour $\tilde{\cal C}$ encircles the cut $[b,a]$ of $\tilde{\cal
H}_H(h)$ and we have changed variables from $h$ to $G$ as above. The contour
integral (\ref{resnontr}) in the complex $G$-plane can be evaluated for
large-$h$, which, by analytical continuation, determines it for all $h$. We can
shrink the contour ${\cal C}_G$ down to an infinitesimal one ${\cal C}_0$
hugging both sides of the cut $[G(b),0]$. In doing so, from (\ref{hG}) it
follows that we pick up contributions from the solution (\ref{sheet1}) and the
extra pole at $G=t$ corresponding to the point $h=\infty$. Thus integrating by
parts we find that (\ref{resnontr}) can be written as
\beq
\tilde{\cal H}_H(h)=\log G_1(h)-\log t+\oint_{{\cal C}_0}\frac{dG}{2\pi
i}~\left\{\frac{\partial}{\partial G}\left[\log
G\log(h-h(G))\right]-\frac{1}{G}\log(h-h(G))\right\}
\label{resshrunk}\eeq

In the contour integral over ${\cal C}_0$ in (\ref{resshrunk}), the total
derivative in $G$ gives the discontinuity across the cut $[G(b),0]$, which is
$\log(h-b)$. The other term there evaluates the $G=0$ limit of $\log(h-h(G))$
determined by (\ref{hG}) and (\ref{sheet1}). Using (\ref{Hsplit}), we then have
\beq
{\cal H}_H(h)=\log G_1(h)+\log\left(\frac{h}{(h-1)t+t^p}\right)
\label{Hsoln}\eeq
Since there is only the single (physical) sheet determined by the singularity
structure of $\e^{{\cal F}(h)}$, we have $\log G_1(h)={\cal F}(h)+{\cal
H}_H(h)$ on the physical sheet, and combined with (\ref{Hsoln}) we arrive at
the expression
\beq
{\cal F}(h)=\log\left(\frac{(h-1)t+t^p}{h}\right)
\label{izlargeN}\eeq
for the large-$N$ limit of the Itzykson-Zuber correlator. As mentioned above,
the fact that $\e^{{\cal F}(h)}$ has only a simple pole of residue $t^p-t$ at
$h=0$ is because there are no other sheets below $G_1(h)$ connected by cuts of
$\e^{{\cal F}(h)}$. This is opposite to the situation that occurs in the
Gaussian case ($A=0$ in (\ref{partfn})), where $\e^{{\cal H}_H(h)}$ has a
simple pole of residue 1 at $h=1$ and there are no upper branches above
$G_1(h)$ connected by cuts of $\e^{{\cal H}_H(h)}$ \cite{kaz1a}. It can be
easily verified, by blowing up the contour $\cal C$, using the asymptotic
boundary condition ${\cal H}_H(h)\sim1/h+{\cal O}(1/h^2)$ for large-$h$ and
computing the residue at $h=\infty$ in (\ref{weightcont}), that
(\ref{izlargeN}) consistently yields the weights (\ref{weightchoice}).
Furthermore, for $p=t=0$, (\ref{izlargeN}) reduces to the solution of
\cite{kaz1a} in the case where only the vertex weight $\tilde t_1^*$ is
non-zero, while for $t=1$ (i.e. $\bar A=\tilde A={\bf1}$), (\ref{izlargeN})
yields ${\cal F}(h)=0$, as expected from its definition.

Notice that for $p\neq1$, ${\cal F}(h)$ here has a logarithmic branch cut
between $h=0$ and $h=\bar t\equiv1-t^{p-1}$, so that strictly speaking the
solution (\ref{izlargeN}) is only valid for $\bar t\leq0$, i.e. $0\leq t\leq1$
for $p\leq0$ and $t\geq a$ for $p>1$ (where its cut doesn't overlap with the
cut $[0,a]$). Outside of this region the analytic structure of the
Itzykson-Zuber correlator can be quite different. For $p=1$, we have ${\cal
F}(h)=\log t$ and the only effect of the Itzykson-Zuber integral in the
saddle-point equation (\ref{saddlepteq}) is to rescale the cosmological
constant as $\lambda^3\to\lambda^3t^2$. This is expected directly from the
original matrix integral (\ref{partfn}), since for $p=1$ the vertex weights can
be represented as traces of the external matrix $A=t\cdot{\bf1}$, while for
$p\neq1$ the $GL(N,{\Bbb C})$ characters can only be defined via the
generalized Schur functions (\ref{schurchar}),(\ref{schurpoly}). For $p\neq1$,
we shall now see that (\ref{izlargeN}) changes the analytic structure of the
large-$N$ solution of the curvature matrix model, but that the $t=1$ physical
characteristics (the pure gravity continuum limit) are unchanged.

The saddle-point equation (\ref{saddlepteq}) with (\ref{izlargeN}) is then
solved by replacing the $t=1$ resolvent ${\cal H}_H(h;\lambda^3,1)$ in
(\ref{resexactHt=1}) by
\beq
{\cal H}_H(h;\lambda^3,t)={\cal H}_H(h;\lambda^3,1)+2\oint_{\cal
C}\frac{ds}{2\pi
i}~\frac{1}{s-h}\sqrt{\frac{(h-a)(h-b)}{(s-a)(s-b)}}~\log\left(
\frac{(s-1)t+t^p}{s}\right)
\label{ressolnt}\eeq
and compressing the closed contour $\cal C$ in (\ref{ressolnt}) to the cut
$[\bar t,0]$. Note that since $\bar t\leq0$ the sign of the square root in
(\ref{ressolnt}) is negative along this cut. Working out the contour
integration in (\ref{ressolnt})  as before then gives
\beq\new{\begin{array}{l}
{\cal H}_H(h;\lambda^3,t)={\cal
H}_H(h;\lambda^3t^{p+1},1)\\~~~~~~~~~~+2\log\left\{\frac{h^2\left(h(a+b)-2ab-
\bar t(2h-a-b)-2\sqrt{(h-a)(h-b)(\bar t-a)(\bar t-b)}\right)}{(h-\bar
t)^2\left(h(a+b)-2ab-2\sqrt{ab(h-a)(h-b)}\right)}\right\}\end{array}}
\label{resexact}\eeq
The endpoints $a,b$ of the support of the spectral distribution function are
found just as before and now the boundary conditions (\ref{bdryeq1}) are
replaced by
\beq
\xi_t=\lambda^6t^{2(p+1)}\eta_t^3~~~~~,~~~~~1-\bar t=t^{p-1}=\eta_t-3\xi_t
\label{bdryeq1t}\eeq
where
\beq
\xi_t=\frac14\left(\sqrt{a-\bar t}-\sqrt{b-\bar
t}\right)^2~~~~~,~~~~~\eta_t=\frac14\left(\sqrt{a-\bar t}+\sqrt{b-\bar
t}\right)^2
\label{xitetatdef}\eeq
and we have assumed that $t\neq0$ \footnote{\baselineskip=12pt It can be easily
seen that for $p=t=0$ the saddle-point equations are not satisfied anywhere.
This simply reflects the fact that it is not possible to close a regular,
planar triangular lattice on the sphere. As discussed in \cite{kaz1,kaz1a}, in
the matrix model formulations one needs to study ``almost" flat planar diagrams
in which positive curvature defects are introduced on the Feynman graphs to
close that triangular lattice on the sphere.}.

The boundary equations (\ref{bdryeq1t}) are identical to those of the $t=1$
model in (\ref{bdryeq1}) with the replacements $\xi\to\bar\xi_t\equiv
t^{1-p}\xi_t$, $\eta\to\bar\eta_t\equiv t^{1-p}\eta_t$ and
$\lambda^3\to\lambda^3t^{p+1}$. The weight average $\langle h\rangle_t$
corresponding to the $1/h^2$ coefficient of the asymptotic expansion of
(\ref{resexact}) is then
\beq
\langle
h\rangle_t=\frac12+t^{2(p-1)}\left(-\frac{\bar\eta_t^2}3+\bar\eta_t-\frac23
\right)
\label{htavg}\eeq
where we have used (\ref{bdryeq1t}). Substituting (\ref{htavg}) into
(\ref{trNX2H}) yields (\ref{trNX2Hexpleta}) with $\eta\to\bar\eta_t$ and an
additional overall factor of $t^{2(p-1)}$ which represents the overall factor
in (\ref{partsumexpl}) (the linear in $\lambda$ term is from the Gaussian
normalization of the partition function). Thus, although the precise analytical
form of the solution is different, the critical behaviour (and also the Wick
expansion) of the curvature matrix model for the choice of vertex weights
(\ref{weightchoice}) with $t\neq1$ is the same as that for $t=1$.

The correlators (\ref{trNX2qH}) will all have a structure similar to those at
$t=1$, as in (\ref{htavg}). The non-trivial analytical structure of the
saddle-point solution for $t\neq1$ is exemplified most in the function
(\ref{Gdef}), which in the present case is
\beq\new{\begin{array}{lll}
G(h;\lambda^3,t)&=&G(h;\lambda^3t^{p+1},1)\\&
&~\times\frac{th^3\left(h(a+b)-2ab-\bar t(2h-a-b)-2\sqrt{(h-a)(h-b)(\bar
t-a)(\bar t-b)}\right)^2}{(h-\bar
t)^3\left(h(a+b)-2ab-2\sqrt{ab(h-a)(h-b)}\right)^2}\\
&=&\frac{\left(h(a+b)-2ab-\bar t(2h-a-b)-2\sqrt{(h-a)(h-b)(\bar t-a)(\bar
t-b)}\right)^2}{\lambda^3t^p(\sqrt a+\sqrt b)^2(a-b)(h-\bar
t)^3\left(h(a+b)-2ab-2\sqrt{ab(h-a)(h-b)}\right)}\end{array}}
\label{Gexpl}\eeq
The inverse function $h(G)$ can be determined from (\ref{Gexpl}) using the
Lagrange inversion formula \cite{di}
\beq
h(G)=G+\sum_{k=1}^\infty\frac1{k!}\left(\frac\partial{\partial
G}\right)^{k-1}\varphi(G)^k
\label{lagrange}\eeq
where the function $\varphi$ is defined from (\ref{Gexpl}) by
\beq
\varphi(h)=h-G(h)
\label{varphidef}\eeq
Comparing with (\ref{hG}) for the choice of vertex weights
(\ref{weightchoice}), we can write down an expression for the generating
function of the dual moments of the triangulation model (or equivalently for
the 4-point and even-even models)
\beq
\sum_{q=1}^\infty\lambda^{-q}\left\langle\frac\tr
N(XA_3)^{3q}\right\rangle_{{\cal
M}_3^*}G^q=G-\frac{t^{p-1}G}{G-t}+\sum_{k=1}^\infty\frac1{k!}\left(\frac
\partial{\partial G}\right)^{k-1}\varphi(G)^k
\label{corrgenfn}\eeq
Because of the complicated structure of the function (\ref{Gexpl}), it does not
seem possible to write down a closed form for this generating function or
systematic expressions for the dual moments. Nonetheless, (\ref{corrgenfn})
does represent a formal solution for the observables of the triangulation
model.

The above critical behaviour, when a non-trivial Itzykson-Zuber correlator is
incorporated into the dynamical triangulation model, is anticipated from
(\ref{critline}) and thus yields a non-trivial verification of the assumptions
that went into the derivation of the large-$N$ limit of the Itzykson-Di
Francesco formula of section 2. The form of the function $G(h;\lambda^3,t)$ in
(\ref{Gexpl}) illustrates the analytical dependence of the saddle-point
solution on the vertex weights $\tilde t_q^*$. It also demonstrates how the
analytical, non-perturbative properties of the random surface sum
(\ref{partsumexpl}) change at $N=\infty$, although the perturbative expansion
of the free energy coincides with (\ref{partsumexpl}) and the physical
continuum limit (\ref{critline}) is unaltered. The discussion of this section
of course also applies to the 4-point and even-even models with the appropriate
redefinitions of coupling constants above, and also to the complex curvature
matrix model where now the incorporation of the Itzykson-Zuber correlator using
the saddle-point equation (\ref{saddlepteqC}) leads to the appropriate
rescaling of the cosmological constant $\lambda^{3/2}$ by $t^{(p+1)/4}$ as
predicted from the graphical arguments of subsection 3.3 (see
(\ref{complexweights})). The above derivation also suggests an approach to
studying phase transitions in the large-$N$ limit of the Itzykson-Zuber model,
as it shows in this explicit example the region where the analytic structure of
${\cal F}(h)$ changes (i.e. $\bar t>0$) and consequently the region wherein a
discontinuous change of the large-$N$ solution appears. This could prove
relevant to various other matrix models where the Itzykson-Zuber integral
appears \cite{mak,semsz}. The saddle-point solution above of the curvature
matrix model can nonetheless be trivially analytically continued to all
$t\in{\Bbb R}$. This is expected since the random surface sum
(\ref{partsumexpl}) is insensitive to a phase transition in the Itzykson-Zuber
integral which appears in the large-$N$ solution of the matrix model only as a
manifestation of the character sum representation of the discretized surface
model.

\section{Complex Saddle Points of the Itzykson-Di Francesco Formula}

The curvature matrix models we have thus far studied have led to a unique,
stable saddle-point solution at large-$N$. From the point of view of the
Itzykson-Di Francesco formula of section 2, there is a crucial reason why this
feature has occured, namely the cancellation of sign factors that appear in
expressions such as (\ref{char3}). The models we have studied have been
arranged so that there is an overall cancellation of such sign factors which
arise from the splitting of the Young tableau weights into the appropriate
congruence classes. When the weights are then assumed to distribute equally the
resulting Vandermonde determinant factors stabilize the saddle-point and lead
to a unique real-valued solution for the free energy and observables of the
random matrix model. In this section we shall briefly discuss the problems
which arise when trying to solve the matrix models when the sign variation
terms in the character expansion formulas do not necessarily cancel each other
out.

The destabilization of the real saddle-point configuration of weights was
pointed out in \cite{kaz1} where it was shown that the configuration is {\it
complex} for the Hermitian one-matrix model with Penner potential \cite{di}
\beq
V_P(XA)=-\log({\bf1}-X)
\label{1matrixpen}\eeq
in (\ref{partfn}). The Itzykson-Di Francesco formula for this matrix model
follows from replacing ${\cal X}_3[h]$ by
$\chi_{\{h\}}(A)=\chi_{\{h\}}({\bf1})\propto\Delta[h]$ in (\ref{diformula}), so
that at $N=\infty$ we have
\beq
Z_P^{(0)}=c_N~\lambda^{-N^2/4}\sum_{h=\{h^e,h^o\}}\Delta[h^o]^2
\Delta[h^e]^2\prod_{i,j=1}^{N/2}
\left(h_i^o-h_j^e\right)\e^{\frac12\sum_ih_i[\log(\frac{\lambda h_i}N)-1]}
\label{partpenner}\eeq
Now there is no problem with the decomposition of weights into the appropriate
congruence classes, but, as we shall see below, the rapid sign changes of the
mixed product over the even and odd weights destabilize the reality of the
saddle-point configuration of Young tableau weights. In the previous models
such mixed product factors did not pose any problem for the solution at
large-$N$ because they appeared in the denominators of the character expansions
and acted to make the more probable configurations of weights those with
identical distributions of even and odd weights, thus stabilizing the
saddle-point at $N=\infty$. In (\ref{partpenner}), however, the mixed product
$\prod_{i,j}(h_i^o-h_j^e)$ appears in the numerator and thus acts to make the
more probable configuration those with different distributions of even and odd
weights. Thus when a symmetric distribution over $h_i^e$ and $h_j^o$ in
(\ref{partpenner}) is assumed, this has the effect of destabilizing the
saddle-point leading to a complex-valued solution.

The matrix model with Penner potential (\ref{1matrixpen}) is equivalent to the
standard Hermitian one-matrix model for pure gravity \cite{fgz}, i.e. that with
potential $\frac14~\tr~X^4$ in (\ref{partfn}). Diagrammatically, a two-to-one
correspondence between planar Feynman graphs of these two matrix models exists
by splitting the 4-point vertices of the $X^4$ model as described in subsection
3.2 to give diagrams of the ``even-log" model with potential
$-\log({\bf1}-X^2)=-\log({\bf1}-X)-\log({\bf1}+X)$ (so that the face centers of
the $X^4$ model are mapped onto the vertices and face centers of the even-log
model). From the point of view of the Itzykson-Di Francesco formula, in the
character expansion (\ref{diformula}) for the $X^4$ model we replace ${\cal
X}_3[h]$ by ${\cal X}_4[h]$. The resulting partition function $Z_4$ is a sum
over mod 4 congruence classes of weights in which the distribution sums for the
classes $\mu=0,2$ and $\mu=1,3$ decouple from each other (so that even and odd
weights completely factorize) and each have the precise form at $N=\infty$ of
the partition function (\ref{partpenner}) \cite{kaz1}, i.e.
$Z_4^{(0)}=(Z_P^{(0)})^2$. This is just the graphical correspondence mentioned
above. Thus the Itzykson-Di Francesco formula at least reproduces correct
graphical equivalences in these cases.

To see how the complex saddle-points arise in the character expansion of the
Penner matrix model above, we assume that even and odd weights distribute
symmetrically in (\ref{partpenner}) and define a distribution function
$\rho_P(h)$ for the $N/2$ weights $h_i^e$. Varying the effective action in
(\ref{partpenner}) for the weights $h^e$ then leads to the large-$N$
saddle-point equation
\beq
\pvint_{\!\!b}^adh'~\frac{\rho_P(h')}{h-h'}=-\frac13\log(\lambda
h)-\log\left(\frac h{h-b}\right)~~~~~,~~~~~h\in[b,a]
\label{pensaddlepteq}\eeq
The corresponding resolvent function ${\cal H}_P(h)$ can be determined by the
contour integration in (\ref{ressoln}) just as before using
(\ref{pensaddlepteq}) and we find after some algebra
\beq
{\cal H}_P(h)=\frac13\log\left\{\frac{(\sqrt a-\sqrt
b)^2(a-b)}{\lambda}\frac{h\left(h+\sqrt{ab}+\sqrt{(h-a)(h-b)}\right)^2}{
\left(h(a+b)-2ab+2\sqrt{ab(h-a)(h-b)}\right)^3}\right\}
\label{respen}\eeq
Expanding (\ref{respen}) for large-$h$ then leads to the boundary conditions
\beq
\xi=\frac{3}{5}\left(\eta-1\right)~~~~~,~~~~~
\left(\frac53\right)^3\lambda^2\eta^5=\left(\eta-1\right)^3
\label{bdryeqpen}\eeq

Consider the structure of the solutions to the boundary conditions
(\ref{bdryeqpen}). Again, the Wick expansion of the original matrix integral is
analytic about $\lambda=0$, and it can be shown that the free energy is
analytic about $\eta(\lambda=0)=1$. We should therefore choose the branches of
the quintic equation in (\ref{bdryeqpen}) which are regular at $\lambda=0$.
There are three solutions which obey this analyticity condition and they are
given by the iterative relations
\beq
\eta_n=1+\frac{5}{3}\omega_3^n\lambda^{2/3}(\eta_n)^{5/3}~~~~~,~~~~~n=0,1,2
\label{recxreg}\eeq
where $\omega_3\in{\Bbb Z}_3$ is a non-trivial cube root of unity. The
remaining two solutions behave near $\lambda=0$ as $\eta\sim\pm\lambda^{-1}$.
The discrete ${\Bbb Z}_3$-symmetry of the regular saddle-point solutions
(\ref{recxreg}) seems to be related to the fact that the Schwinger-Dyson field
equations of this matrix model determine the function $G_P(h)=\e^{{\cal
H}_P(h)}$ as the solution of a third-order algebraic equation \cite{kaz1}
\beq
\lambda h^3G_P^3-\lambda
h^2(1+h)G_P^2+\left[\frac89-h+\frac1{648\lambda}\left(1-\sqrt{1-12\lambda}
\right)(1-12\lambda)\right]hG_P+h^2=0
\eeq

Initially, the endpoints $a,b$ of the support of the spectral density lie on
the positive real axis, so that one expects that only the real branch $\eta_0$
 is a valid solution of the matrix model. However, the perturbative
expansion parameter of the free energy $S_P(\eta_0)$ would then be
$\lambda^{2/3}$. It is easy to see by a Wick expansion of the original matrix
integral that the genus zero expansion parameter is in fact $\lambda^2$.
 Furthermore, one
can analyse the analytic properties of the solutions to the quintic boundary
equation in (\ref{bdryeqpen}), and even determine the critical point
$\lambda_c$ which in this case is the point where the two real and positive
roots coalesce. For $\lambda>\lambda_c$ all three roots which are
analytic about $\lambda=0$ become complex-valued.
This critical behaviour is similar
to that discussed for the cubic boundary equation (\ref{bdryeq1}) in subsection
3.4, and so apparently lies in the same universality class as the earlier
models.
However, the
critical value $\lambda_c$ determined this way does not agree with known
results \cite{fgz,akm}, so that the free energy $S_P$ determined from the
character expansion does not count the Feynman graphs of the Penner
model  correctly.

The structure of complex saddle-points arises  in many other matrix models. For
example,
 for the quartic-type Penner
potential
\beq
V_P^{(4)}(XA)=-\log({\bf1}-X^4)
\label{pen4pot}\eeq
 the boundary conditions determining the
endpoints $a,b$ of the support of the distribution function are
\beq
\xi=\frac{3}{7}\left(\eta-1\right)~~~~~,~~~~~
\left(\frac73\right)^3\lambda^4\eta^7=\left(\eta-1\right)^3
\label{bdryC}\eeq
Again, there are three regular solutions of (\ref{bdryC}) at $\lambda=0$
and the real root leads to an expansion in $\lambda^{4/3}$, whereas
the Wick expansion can be explicitly carried out and one
finds that the perturbative expansion parameter is $\lambda^4$.

All the models we have studied for which the matrix model saddle point
does not reproduce the graphical expansion share the feature that the
constraint of regularity at $\lambda=0$ yields
multiple  solutions of the endpoint equations. Choosing the real root
(or any other single root) leads to the wrong solution
of the matrix model.  Thus it appears that  the saddle-point
should  be  defined by some kind of
 analytical continuation  that extends the support of the spectral density
$\rho_P$
 into the complex plane and takes proper account of the multiple root
structure.
It would be interesting to resolve these problems and determine the general
technique for dealing with such complex saddle-points. It would also be
interesting to discover any connection between  this saddle-point
destabilization
and the
well-known occurence of (single-branch) complex saddle points for the
eigenvalue distributions in generalized Penner models \cite{akm}.

\section{Conclusions}

In this paper we have shown that the character expansion techniques developed
in \cite{kaz1,kaz1a} can be applied to odd potentials. We have demonstrated
that the splitting of weights into congruence classes other than those of the
even representations of $GL(N,{\Bbb C})$ leads to a proper solution of the
matrix model, provided that one writes the weight distribution that appears in
the character expansion over the appropriate congruence elements. The
Itzykson-Di Francesco formula then correctly reproduces relations between
different models.

{}From a mathematical perspective, the results of this paper raise some
questions concerning the large-$N$ limit of the Itzykson-Di Francesco formula.
For instance, a random surface model with a discrete ${\Bbb Z}_p$-symmetry
corresponding to, say, a $p$-colouring of its graphs will be described by a
curvature matrix model with the ${\Bbb Z}_p$-symmetry $A\to\omega_pA$,
$\omega_p\in{\Bbb Z}_p$. This symmetry will be reflected in the Itzykson-Di
Francesco expansion as a localization of the group character sum onto mod $p$
congruence class representations of $GL(N,{\Bbb C})$. The appropriate solution
of the model at $N=\infty$ will then involve resumming the Young tableau
weights over the mod $p$ congruence classes and assuming that the even-odd
decomposition factorizes symmetrically over these classes. However, it is not
immediately clear why such a symmetry assumption of the character expansion at
large-$N$ gives the appropriate solution of the discretized planar surface
theory (although a mapping onto a complex matrix model indicates how this
should work). At this stage there seems to be a mysterious ``hidden" symmetry
at play which makes the large-$N$ group theoretical approach to solving these
random surface models work. Furthermore, other intriguing features of the
Itzykson-Di Francesco formula, such as the appearence of phase transitions in
the Itzykson-Zuber integral represented through the large-$N$ limit of
generalized Schur functions, are purely a result of the character expansion
representation and correctly conspire to yield the proper solution of the
random surface model. It would be interesting to put all of these features into
some systematic framework for dealing with curvature matrix models in general.

\begin{figure}
\unitlength=0.90mm
\linethickness{0.4pt}
\begin{picture}(150.00,70.00)(0,10)
\small
\put(70.00,15.00){\line(1,0){50}}
\put(70.00,15.00){\line(1,2){25}}
\put(120.00,15.00){\line(-1,2){25}}
\put(70.00,15.00){\line(-1,-1){4}}
\put(120.00,15.00){\line(1,-1){4}}
\put(95.00,65.00){\line(0,1){4}}
\put(94.00,15.00){\line(-1,2){12}}
\put(96.00,15.00){\line(1,2){12}}
\put(83.00,41.00){\line(1,0){24}}
\put(81.00,15.00){\line(-1,2){5.5}}
\put(83.00,15.00){\line(1,2){5.5}}
\put(76.50,28.00){\line(1,0){11}}
\put(107.00,15.00){\line(-1,2){5.5}}
\put(109.00,15.00){\line(1,2){5.5}}
\put(102.50,28.00){\line(1,0){11}}
\put(94.00,41.00){\line(-1,2){5.5}}
\put(96.00,41.00){\line(1,2){5.5}}
\put(89.50,54.00){\line(1,0){11}}
\put(74.50,15.00){\line(-1,2){2.25}}
\put(76.50,15.00){\line(1,2){2.25}}
\put(73.25,21.50){\line(1,0){4.5}}
\put(87.50,15.00){\line(-1,2){2.25}}
\put(89.50,15.00){\line(1,2){2.25}}
\put(86.25,21.50){\line(1,0){4.5}}
\put(100.50,15.00){\line(-1,2){2.25}}
\put(102.50,15.00){\line(1,2){2.25}}
\put(99.25,21.50){\line(1,0){4.5}}
\put(113.50,15.00){\line(-1,2){2.25}}
\put(115.50,15.00){\line(1,2){2.25}}
\put(112.25,21.50){\line(1,0){4.5}}
\put(81.00,28.00){\line(-1,2){2.25}}
\put(83.00,28.00){\line(1,2){2.25}}
\put(79.75,34.50){\line(1,0){4.5}}
\put(107.00,28.00){\line(-1,2){2.25}}
\put(109.00,28.00){\line(1,2){2.25}}
\put(105.75,34.50){\line(1,0){4.5}}
\put(87.50,41.00){\line(-1,2){2.25}}
\put(89.50,41.00){\line(1,2){2.25}}
\put(86.25,47.50){\line(1,0){4.5}}
\put(100.50,41.00){\line(-1,2){2.25}}
\put(102.50,41.00){\line(1,2){2.25}}
\put(99.25,47.50){\line(1,0){4.5}}
\put(94.00,54.00){\line(-1,2){2.25}}
\put(96.00,54.00){\line(1,2){2.25}}
\put(92.75,60.50){\line(1,0){4.5}}
\end{picture}
\begin{description}
\small
\baselineskip=12pt
\item[Figure 5:] An example of a regular fractal-type graph that appears in the
dynamical triangulation.
\end{description}
\end{figure}

{}From a physical point of view, there are many random surface models which are
best dealt with using the dynamical triangulations studied in this paper, and
it would be interesting to exploit the relationships with the even coordination
number models to study the properties of these theories. For instance, one
sub-ensemble of the random surface sum (\ref{partgraphs}) is the collection of
regular fractal-like graphs (Fig. 5) which were shown in \cite{hw} to dominate,
in the high-temperature limit, the surface sum for two-dimensional quantum
gravity coupled to a large number of Ising spins when restricted to
two-particle irreducible Feynman diagrams. These fractal graphs can be
characterized by 3-point graphs $G_3$ where only the dual coordination numbers
\beq
q^*_k=3\cdot2^k~~~~~~,~~~~~~k\geq0
\label{fractalq}\eeq
are permitted. Here $k$ is the order of the fractal construction obtained
inductively by replacing each 3-point vertex of an order $k-1$ fractal graph
with a triangle\footnote{\baselineskip=12pt Note that the total number of
triangle sides along each outer side of the fractal-like structure of a single,
order $k+1$ fractal graph is $2^k-1$. Thus to be able to close the set of
3-point graphs of dual coordination numbers (\ref{fractalq}) on a spherical
topology (corresponding to $N=\infty$ in the matrix model), one needs to glue
an order $k$ and an order $k+1$ fractal graph together along their three
external corner legs (see Fig. 5) so that the valence of the dual vertices of
the faces joining the two fractal structures will coincide with
(\ref{fractalq}).}. The ensemble of fractal graphs corresponds to a branched
polymer phase of two-dimensional quantum gravity \cite{polymer}. We have shown
that when the $3m$-sided polygons in (\ref{partgraphs}) are weighted with a
power law variation with the number of sides, the continuum limit of the model
lies in the pure gravitational phase. The curvature matrix model (\ref{partfn})
with the dual vertex weights arranged as discussed in this paper can thus serve
as an explicit model for the transition from a theory of pure random surfaces
(associated with central charge $D<1$) to a model of branched polymers
(associated with $D\geq1$). This might help in locating the critical dimension
$D_c\geq1$ where the precise branched polymer transition takes place.

\end{document}